\documentclass[fleqn,usenatbib]{mnras}

\usepackage{newtxtext,newtxmath}
\usepackage{threeparttable}

\usepackage{ae,aecompl}

\usepackage{graphicx}	
\usepackage{amsmath}	
\usepackage{amssymb}	
\usepackage{multicol}
\usepackage{hyperref}

\title[GMRT 610~MHz survey of ELAIS N1]{A Wide-area GMRT 610~MHz survey of ELAIS N1 field}

\author[Ishwara-Chandra et~al.]{C.~H.\ Ishwara-Chandra,$^{1,2}$\thanks{E-mail:
   ishwar@ncra.tifr.res.in}
  A.~R.\, Taylor,$^{2,3}$  D.~A.\ Green,$^4$ J.~M.\ Stil,$^{5}$\newauthor
  M.\ Vaccari,$^{3,6}$, E.~F.\, Ocran$^2$\\
$^1$National Centre for Radio Astrophysics, Tata Institute of Fundamental 
Research, Pune University Campus, Pune 411007, India \\
$^2$Inter-university Institute for Data Intensive Astronomy, Department of Astronomy, University of Cape Town, \\ 
7701 Rondebosch, Cape Town, South Africa\\
$^3$Inter-university Institute for Data Intensive Astronomy, Department of Physics and Astronomy, \\
University of the Western Cape, 7535 Bellville, Cape Town, South Africa\\
$^4$Cavendish Laboratory, University of Cambridge, Cambridge, UK\\
$^5$Department of Physics and Astronomy, University of Calgary, Calgary, Canada\\
$^6$INAF - Istituto di Radioastronomia, via Gobetti 101, 40129 Bologna, Italy\\
}
\date{Accepted XXX. Received YYY; in original form ZZZ}

\pubyear{2015}

\begin{document}
\label{firstpage}
\pagerange{\pageref{firstpage}--\pageref{lastpage}}
\maketitle

\begin{abstract}
In this paper we present a wide-area 610 MHz survey of the ELAIS\,N1  field with the GMRT, covering an area of 12.8 deg$^2$ at a resolution of 6 arcsec and with an rms noise of $\sim 40$ $\mu$Jy beam$^{-1}$. This is equivalent to $\sim 20$ $\mu$Jy beam$^{-1}$ rms noise at 1.4 GHz for a spectral index of $-0.75$.  The primary goal of the survey was to study the polarised sky at sub-mJy flux densities at $<$ GHz frequencies. In addition, a range of other science goals, such as investigations in to the nature of the low-frequency $\mu$Jy source populations and alignments of radio jets.
A total of 6,400 sources were found in this region, the vast
majority of them compact. The  sample jointly detected by GMRT at 610 MHz and by VLA FIRST at 1.4\,GHz has a median spectral index of $-0.85 \pm 0.05$ and a median 610 MHz flux density
of 4.5 mJy. This region has a wealth of ancillary data which is useful to characterize the detected sources. The multi-wavelength cross matching resulted optical/IR counterparts to 
$\sim 90$ per~cent of the radio sources, with a significant fraction having at least photometric redshift. Due to the improved sensitivity of this survey over preceding ones, we have discovered six giant radio sources (GRS), with three of them at $z \sim 1$ or higher. This implies that the population of GRS may be more abundant and common than known to date and if true this has implications for the luminosity function and the evolution of radio sources. We have also identified several candidate extended relic sources.
\end{abstract}

\begin{keywords}
  surveys -- galaxies: active -- radio continuum: galaxies
\end{keywords}



\section{Introduction}

Deep and wide-area radio surveys are powerful tools to study a range of
source populations. Since radio waves are not obscured by dust, deep
radio surveys are also ideal to search for distant objects which are
otherwise dust-obscured and thus not detected in the
Ultra-violet(UV)/Optical.  As can be seen from major radio surveys, the
radio sky is homogeneous as compared to the optical sky
\citep{2007ASPC..380..189C}. A radio survey will thus help to unveil
otherwise obscured galaxy populations.

The radio sources broadly fall in two categories -- normal  and active
galaxies. In the first case, the radio emission is due to
secondary processes related to star formation such as from supernova
remnants and \ion{H}{ii} regions. Such close relation between the radio
emission and star formation in normal galaxies is evident from the tight
radio -- infrared correlation \citep{1992ARA&A..30..575C}.  The radio
band provides sensitive and independent estimate of star formation rate
which is particularly useful where the galaxies are affected by dust
obscuration and estimating star formation rate using   the optical band could
be affected by several observational biases. In the second case, the radio
emission is due to relativistic electrons in lobes which forms the
termination region of jets powered by super-massive black holes at the
center of massive ellipticals. Such galaxies where there is  a
super-massive black hole at the centre which forms an accretion disk  are
known as Active Galaxies or Active Galactic Nuclei (AGN). AGNs are known
to emit powerful radiation across the electromagnetic spectrum.  A small
fraction of them ($\sim 10$ per~cent) emit powerful radio
emission fueled by jets which  forms hotspots and lobes on either side
of the host galaxy. The size of such radio sources can be as large as
several mega parsecs.  The strength of radio emission in these 
AGNs can be several orders of magnitude more than the emission in
optical band and is referred to as `radio loud' AGNs
\citep{1989AJ.....98.1195K}.  These radio loud AGNs will significantly
deviate from the radio--FIR correlation. For this reason, the radio -
FIR correlation is a powerful way to disentangle populations of radio
sources where the emission is due to stellar origin and that due to AGN
process. It has been widely accepted now that star formation is
regulated or suppressed by the feedback from AGNs in galaxies
\citep{2012ARA&A..50..455F}. A radio source where the radio emission is
due to AGN process, naturally implies that the galaxy hosts a
supermassive black hole. Therefore detecting number of radio sources
where the radio emission is due to AGN process is important to
understand the evolution of the host galaxy from the context of AGN
feedback.

In a given radio survey where both normal galaxies and radio loud AGNs
are present it is possible to classify radio sources and measure
their distances through their optical counterparts. In early radio
surveys such as 3CRR and 4C, most of the radio sources were radio loud
AGNs.  However at sub-mJy sensitivity, normal galaxies start contributing
to the source population. Early deep radio surveys have helped to
classify the strong radio sources ($\ga$ a few mJy) into that powered
by AGNs while the faint radio sources are predominantly starbursts and
normal galaxies \citep{1985ApJ...289..494W}.  Since then, there have
been several very deep radio surveys where the fraction of radio sources
powered by AGN and starbursts were clearly identified
\citep{2011ApJ...740...20P, 2013MNRAS.436.3759B} using the tight
relation between far infra-red and radio band and X-Ray data. This also
helped to identify a third category of radio source where radio emission
from radio quiet AGNs. As on today, it is widely accepted that at sub-mJy
flux levels at 1.4 GHz, the normal galaxies dominate the radio survey
where as stronger radio sources are mostly radio loud AGNs. For a
statistically meaningful studies about fraction of radio loud AGNs, star
forming galaxies, radio quiet AGN and other atypical radio sources deep
as well as wide area surveys are required. Such surveys also have been
proved useful for serendipitous discoveries of rare class of objects
like relics \citep{2016A&A...585A..29B}.

The radio emission is dominated by the synchrotron processes with a typical
spectral index, $\alpha$, of $-$0.7 (flux density, $S$, scaling with
frequency $\nu$ as $S \propto \nu^{\alpha}$).
This means, for a given depth, there will be far more sources
at lower radio frequencies as compared to at higher radio frequencies.
In addition, a subset of radio sources preferentially show steep radio
spectra with spectral index  $\alpha < -1$, hence low frequency surveys
are better suited to study these objects. One such population is
high-redshift  radio galaxies which tend to exhibit steeper radio
spectra and this has been exploited to discover several high-redshift
radio galaxies \citep{1979A&A....80...13B, 2010MNRAS.405..436I, 2018MNRAS.480.2733S}.
Therefore for discovering high-redshift radio galaxies, the radio
surveys at low radio frequencies ($<1$ GHz) are preferred
\citep{2012MNRAS.420.2644K}. Also, for GHz peaked spectrum sources (GPS)
where the radio emission peaks at 1 to 2 GHz in rest frame, the peak is
redshifted to  a few hundred MHz for high redshift GPS sources
\citep{2017ApJ...836..174C}. In addition to such sources where low
frequency will pick them up preferentially, normal radio sources will also exhibit steeper spectral index if the radio jet powering the radio lobes is switched off where the radio emitting relativistic electrons are decaying.  Such
steep spectral indices are characteristics of relic radio emission
\citep{2016A&A...585A..29B}. It has been widely believed that every AGN
is likely to go through a radio loud phase which implies that radio
emission from the previous epoch of activity will exhibit a steep radio
spectrum and will be seen only in low frequency radio surveys. Hitherto,
examples of such episodic AGNs are very scarce owing to lack of deep and
wide radio surveys particularly at frequencies below 1 GHz. To
summarize, low frequency radio surveys are  useful to discover
high-redshift radio galaxies, high-redshift GPS sources and relic radio
emission due to episodic AGN activity which are often missed at GHz
frequencies. Most of the deep radio surveys in the past are carried out
at 1.4 GHz or higher radio frequencies which is biased against steep
spectrum objects and very few deep surveys exists at $<1$ GHz with wide
area and rms noise of $<0.1$ mJy beam$^{-1}$
\citep{2007A&A...463..519B}.

Here we present one of the deepest and wide area survey of ELAIS\,N1
region (European Large Area ISO Survey North 1)
\citep{2000MNRAS.316..749O}  with the GMRT at 610 MHz, covering $\sim
12.8$ deg$^2$ at a resolution of 6 arcsec and with an rms noise of
$\sim 40$ $\mu$Jy beam$^{-1}$. These observations are deeper than
previous 610 MHz observations with the GMRT \citep{2008MNRAS38375G}. For
a spectral index of $-$0.75, the present observations correspond to an rms
noise of $\sim 20$ $\mu$Jy beam$^{-1}$ at 1.4 GHz which is
substantially deeper than VLA FIRST survey \citep{1995ApJ...450..559B}. Recently deeper but a smaller area of this field was observed at 400 MHz with the upgraded GMRT  \citep{2019MNRAS.490..243C}.
This region was chosen for several reasons such as availability of
plenty of multi-wavelength data.

The paper is arranged as follows. The observations and data analysis is described in Section 2. The source catalogue is presented in Section 3. The
multi-wavelength information is given in Section 4 and notes on a few
individual  extended objects are given in Section 5.
%
%

\section{Observations and data reduction}

The ELAIS\,N1 field was originally chosen for deep extragalactic
observations with the Infrared Space Observatory (ISO) due to its low
infrared background \citep{RowanRobinson2004,Vaccari2005}. Since then,
it has gone on to become one of the best-studied 1--10 deg$^2$
extragalactic fields. The ELAIS\,N1 region, covering $\sim 10$ deg$^2$
was observed at 610 MHz using the Giant Metrewave Radio Telescope
(GMRT). The GMRT consists of 30 antennas spread over $\sim 25$ km$^2$
area situated in western India \citep{1991CuSc...60...95S}. A total of 51 pointings were employed to cover this area.  The observations for most
of the pointings were carried out in 2011 and 2012 and a few remaining
pointings were observed in December 2017.  The data were recorded using the GMRT Software Backend (GSB) with 32 MHz bandwidth and 256 spectral
channels. Typically, each pointing was split over multiple days to
optimise  $u$-$v$ coverage.   The primary calibrator 3C286 was observed on all
days (additionally 3C48 on a few days) both at the start and at the end
of each observing session for flux and bandpass calibration. The flux
density was set using the Perley--Butler 2013 scale
\citep{2013ApJS..204...19P}, with 22.179 Jy for 3C286 and 31.238 Jy for
3C48 at the channel 0 frequency of 596 MHz. Each pointing was observed
for about 30 minutes and then a phase calibrator was observed for 5
minutes for gain calibrations. Total  on-source time on each
pointing was typically $\sim 2.5$ hours.

\begin{figure}
\centerline{\includegraphics[width=0.95\linewidth]{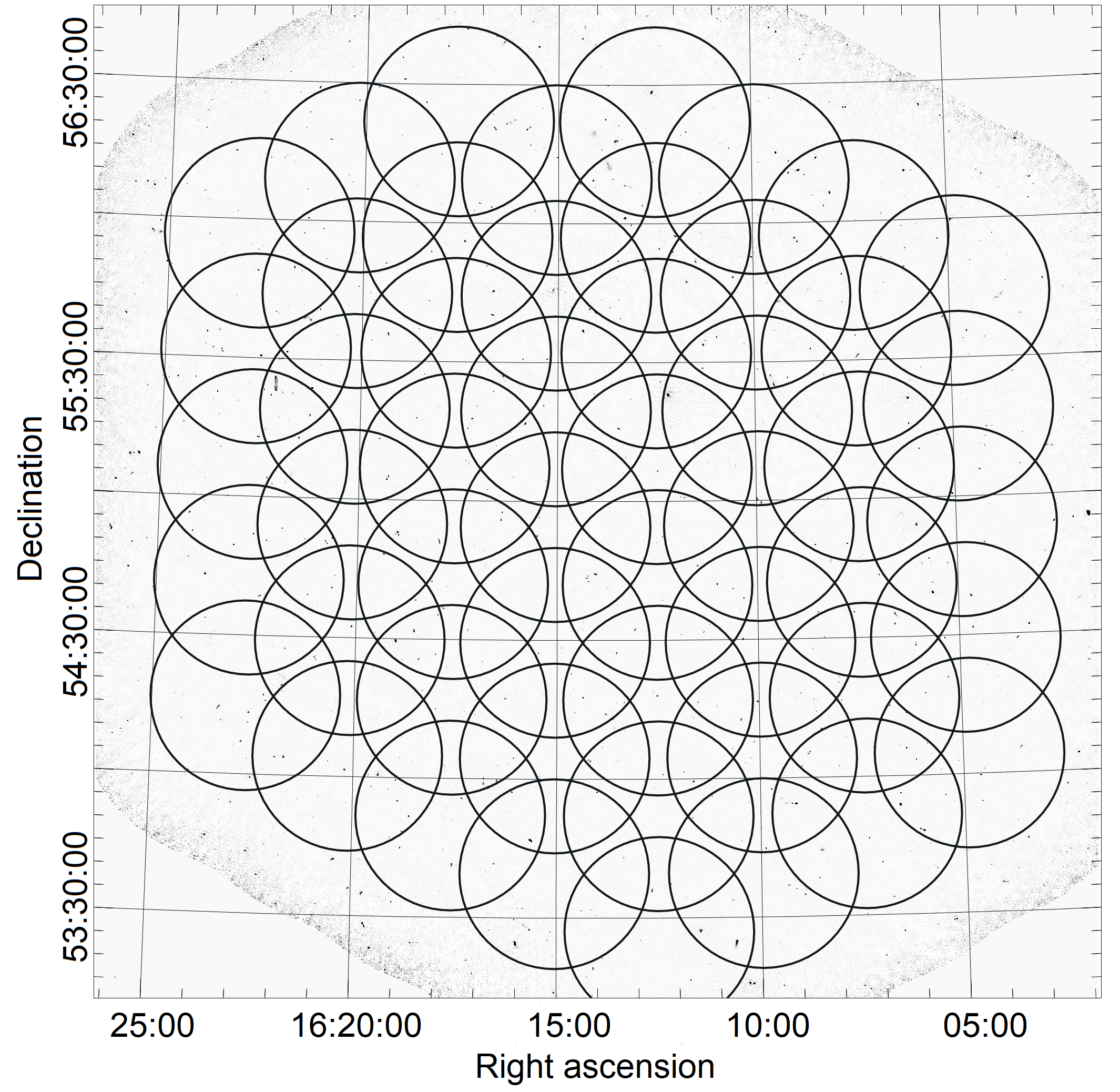}}
\caption{The array of 51 pointing centres overlaid on the ELAIS\,N1
610\,MHz mosaic. Each pointing is represented by a circle with diameter
equal to the FWHM of the primary beam of $41$ arcmin. The pointings are
separated by $25$ arcmin. The area shown is $3.6 \times 3.6$ deg$^2$ and
the imaged portion covers 12.8 deg$^2$.}
\label{fig:geometry}
\end{figure}

\begin{figure*}
\centerline{\includegraphics[width=\linewidth]{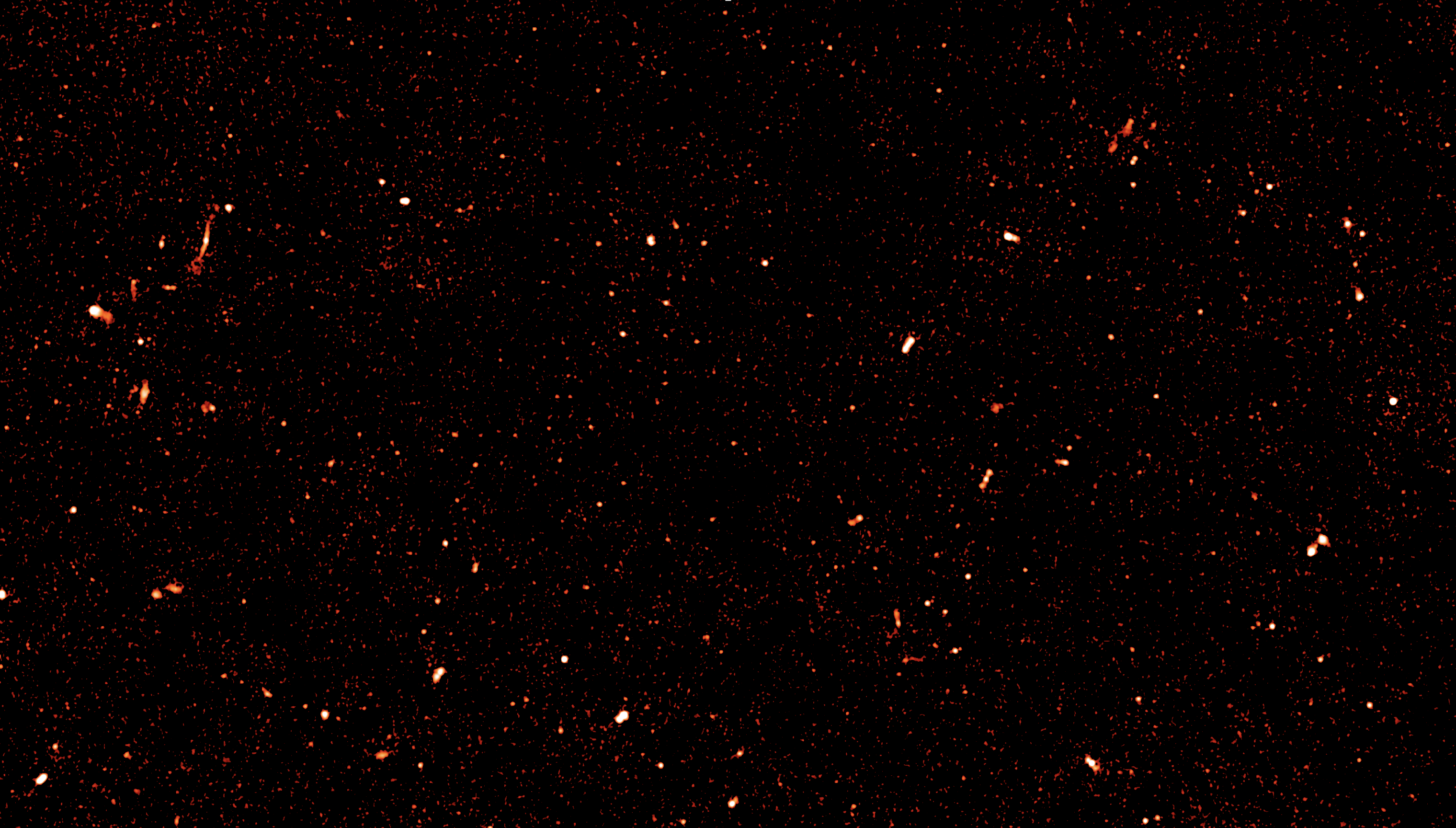}}
\caption{A $1.5\times 0.5$ deg$^2$ section of the ELAIS\,N1 610\,MHz
image centred at $\alpha = 16^{\rm h}\,11^{\rm m}\,12^{\rm s}$ and
$\delta = 54^{\circ}\,10'\,15''$. The resolution is 6 arcsec and the rms
noise in this region of the mosaic (away from bright sources) is $\sim
40$ $\mu$Jy beam$^{-1}$. The intensity scale is logarithmic between 60
$\mu$Jy beam$^{-1}$ and 2 mJy beam$^{-1}$.}
\label{fig:en1w}
\end{figure*}

The data analysis was carried out on the data intensive cloud at the
Inter-University Institute for Data Intensive Astronomy (IDIA), using a
{\sc casa} (Common Astronomy Software Applications) based pipeline.
Initial flagging was done using the casa task {\tt flagdata}.  First, outliers were removed using the `clip' mode and the data was flagged using the `tfcrop' mode. The delay, bandpass and gain calibration were carried out on flagged data. Post-calibration, the data were flagged again, using `clip', `tfcrop' and `rflag' modules.
The data was re-calibrated after removing the initial calibration. The central $\sim$ 90\% band was used for channel averaging with
post-averaging channel width of 0.78 MHz to keep the bandwidth smearing
negligible. On the science target, the total flagged data was $<10$
percent in most fields. Split files from each pointing (if observed on
different epochs), were combined using {\tt concat} before imaging. The
imaging was carried out using the task {\tt tclean} in {\sc casa}. Four
rounds of phase-only self-cal with solution intervals of 8, 4, 2 and 1
minutes and then 5 rounds of amplitude and phase self-cal with solution
intervals of 8, 4, 2, 1 and 1 minutes was carried out on each pointing. 
The entire pipeline for calibration and
imaging was carried out in nearly automatic manner. The input to the
pipeline is the multi-source FITS file and other book keeping
information regarding the calibrators, target name, channels to average
and other imaging and calibration parameters. For convenience, the
pipeline was divided into two parts as calibration and imaging. In the
calibration part, the pipeline starts with importing the multi-source
fits file and ends after split which produces ready-to-image calibrated visibilities. After this, common pointings observed on multiple days were
concatenated (outside the pipeline) which forms the input to the imaging
pipeline. A field of view of about 2.3 times the FWHM (Full Width at
Half Maximum of the primary beam) was imaged to cover up to the first sidelobe. In the first round of imaging, 
the number of {\tt clean} iterations was limited to 5000. A flux
threshold of  approximately 20 times the thermal noise was kept for the
first round of imaging. In subsequent self-calibration and imaging
rounds, the number of {\tt clean} iterations was doubled with moderate
lowering of the {\tt clean} threshold. Also after the first round of
phase-only self-calibration, flagging based on residuals  (corrected\_data $-$ model\_data) was employed in
2nd, 3rd and fourth round of phase-only self-calibration. We have
observed that the residual based flagging performs better than   
flagging on the corrected\_data column.

After four rounds of phase-only self-calibration, the pipeline enters
the amplitude-and-phase self-calibration loop. Here also, from 2nd round
onwards, the data were flagged based on residuals. The total flagged data
was $<30$ per cent in most fields.  The final solution interval for
self-calibration was limited to 1 minute in the fourth and fifth round.
This resulted in significant improvement in image fidelity. The rms
noise on the final individual images  was $\sim 50$ $\mu$Jy
beam$^{-1}$  before the primary beam correction in most of
the images. Moderately higher noise was observed in fields with radio
sources stronger than $\sim 100$ mJy.

The 51 pointings were mosaicked to create an image of EN1 covering an
area of 12.8 deg$^2$.  The mosaic geometry is illustrated in
Figure~\ref{fig:geometry}.  Due to the different $u$-$v$ coverage for
each pointing, the FWHM of the synthesized beam varies between 4.5 and 6
arcsec.  Before mosaicing the image from each field  was smoothed to a
circular Gaussian beam with FWHM of 6 arcsec. The mosaic, including data
from each field up to 20 per cent level of the primary beam, was carried
out in {\sc AIPS} using the python script {\tt make\_mosaic} (Intema,
private communication).  The details of weighting schemes employed in
the mosaic process is described in \citet{2017A&A...598A..78I}. The
final rms in the total intensity mosaic image is $\sim 40$ $\mu$Jy
beam$^{-1}$. The resulting mosaic is about $3.6 \times 3.6$ deg$^2$ and
has total imaged solid angle of 12.8 deg$^2$ (Figure
\ref{fig:geometry}). A sample sub-region of the mosaic image is shown in
Figure~\ref{fig:en1w}.

\section{Source Catalogue}

\begin{table*}
\small
\caption{A small portion of the GMRT ELAIS\,N1 Wide 610\,MHz Source Catalogue. Full catalogue is available online.}
\label{tab:catalogue}
\begin{tabular}{ccccccrrrc} \hline
    (1)      &  (2)     &  (3)    & (4)   &  (5)   & (6)   & (7)  & (8)   &   (9)   &  (10)  \\
Name & RA(J2000) & DEC(J2000) &  $S_{\rm I}$ &   $S_{\rm p}$ &   RMS & \hfill $\theta_{\rm maj}$ \hfill &  \hfill $\theta_{\rm min}$ \hfill & PA \hfill& Code \\    
& deg$\pm$arcsec & deg$\pm$arcsec &  mJy &  mJy  & mJy &   \hfill arcsec \hfill &  \hfill arcsec \hfill & deg\hfill& \\
 \hline
J161300+532637 & 243.24964 $\pm$ 0.46 & 53.44416 $\pm$ 0.50  & 0.55 $\pm$  0.10  & 0.39  $\pm$ 0.06& 0.058 &  7.3  $\pm$  1.2 &6.9 $\pm$ 1.1& 24.7 $\pm$ 105.0 & S \\
J161300+554834 & 243.24992 $\pm$ 0.35 & 55.80993 $\pm$ 0.34  & 0.27 $\pm$  0.08  & 0.31  $\pm$ 0.04& 0.044 &  6.1  $\pm$  0.9 &5.2 $\pm$ 0.7& 47.1 $\pm$  35.9 & S \\
J161300+543542 & 243.25302 $\pm$ 0.28 & 54.59539 $\pm$ 0.32  & 0.34 $\pm$  0.09  & 0.38  $\pm$ 0.05& 0.049 &  6.0  $\pm$  0.8 &5.4 $\pm$ 0.6&153.5 $\pm$  52.9 & S \\
J161300+544502 & 243.25312 $\pm$ 0.59 & 54.75106 $\pm$ 0.54  & 0.36 $\pm$  0.08  & 0.27  $\pm$ 0.05& 0.050 &  7.4  $\pm$  1.5 &6.4 $\pm$ 1.1& 54.7 $\pm$  59.0 & S \\
J161300+555238 & 243.25336 $\pm$ 0.73 & 55.87776 $\pm$ 1.95  & 4.44 $\pm$  0.20  & 1.25  $\pm$ 0.14& 0.082 & 23.6  $\pm$  4.9 &5.7 $\pm$ 0.6& 70.5 $\pm$   8.9 & M \\
J161300+534130 & 243.25364 $\pm$ 0.06 & 53.69203 $\pm$ 0.06  & 1.83 $\pm$  0.08  & 1.68  $\pm$ 0.06& 0.038 &  6.4  $\pm$  0.1 &6.1 $\pm$ 0.1& 68.6 $\pm$  22.6 & S \\
J161301+541001 & 243.25443 $\pm$ 0.12 & 54.16748 $\pm$ 0.14  & 0.97 $\pm$  0.07  & 0.83  $\pm$ 0.05& 0.038 &  6.7  $\pm$  0.3 &6.3 $\pm$ 0.3&178.4 $\pm$  29.5 & S \\
J161301+535251 & 243.25504 $\pm$ 0.22 & 53.88111 $\pm$ 0.24  & 0.31 $\pm$  0.06  & 0.36  $\pm$ 0.04& 0.036 &  5.7  $\pm$  0.6 &5.4 $\pm$ 0.5&169.9 $\pm$  89.5 & S \\
J161301+533521 & 243.25613 $\pm$ 0.42 & 53.58959 $\pm$ 0.26  & 0.33 $\pm$  0.07  & 0.32  $\pm$ 0.04& 0.041 &  7.0  $\pm$  1.0 &5.3 $\pm$ 0.6&103.8 $\pm$  20.9 & S \\
J161301+533851 & 243.25670 $\pm$ 0.30 & 53.64803 $\pm$ 0.89  & 0.46 $\pm$  0.10  & 0.24  $\pm$ 0.04& 0.042 &  8.2  $\pm$  2.2 &2.9 $\pm$ 0.2& 72.2 $\pm$   7.0 & M \\
 \hline
\end{tabular}
\end{table*}

The source catalogue was created using PyBDSF \citep{PyBDSF2015}. In
order to minimise the possibility of spurious candidates near bright
sources and at the edge, we have adopted advanced options. Near the
bright sources we have adopted a smaller rms box size of $33 \times 33$
pixel$^2$ sliding by 11 pixels. In the rest of the image, the rms box
size of $40 \times 40$ pixel$^2$ sliding by 13 pixels was used.
This scheme resulted in least number of spurious sources near bright
sources on visual inspection, as compared to other box sizes that were
tried. For source detection, a 5$\sigma$ threshold for peak
pixel and 3$\sigma$ cutoff for island boundary was used. We had invoked
the option to group overlapping Gaussians into a single source.  With these 
criteria, a total of 6,474 sources were catalogued. All the sources were
visually inspected and 74 sources which were likely to be noise peaks in
the high noise region near bright sources were removed.

\begin{figure}
\centerline{\includegraphics[width=2.15in,angle=270]{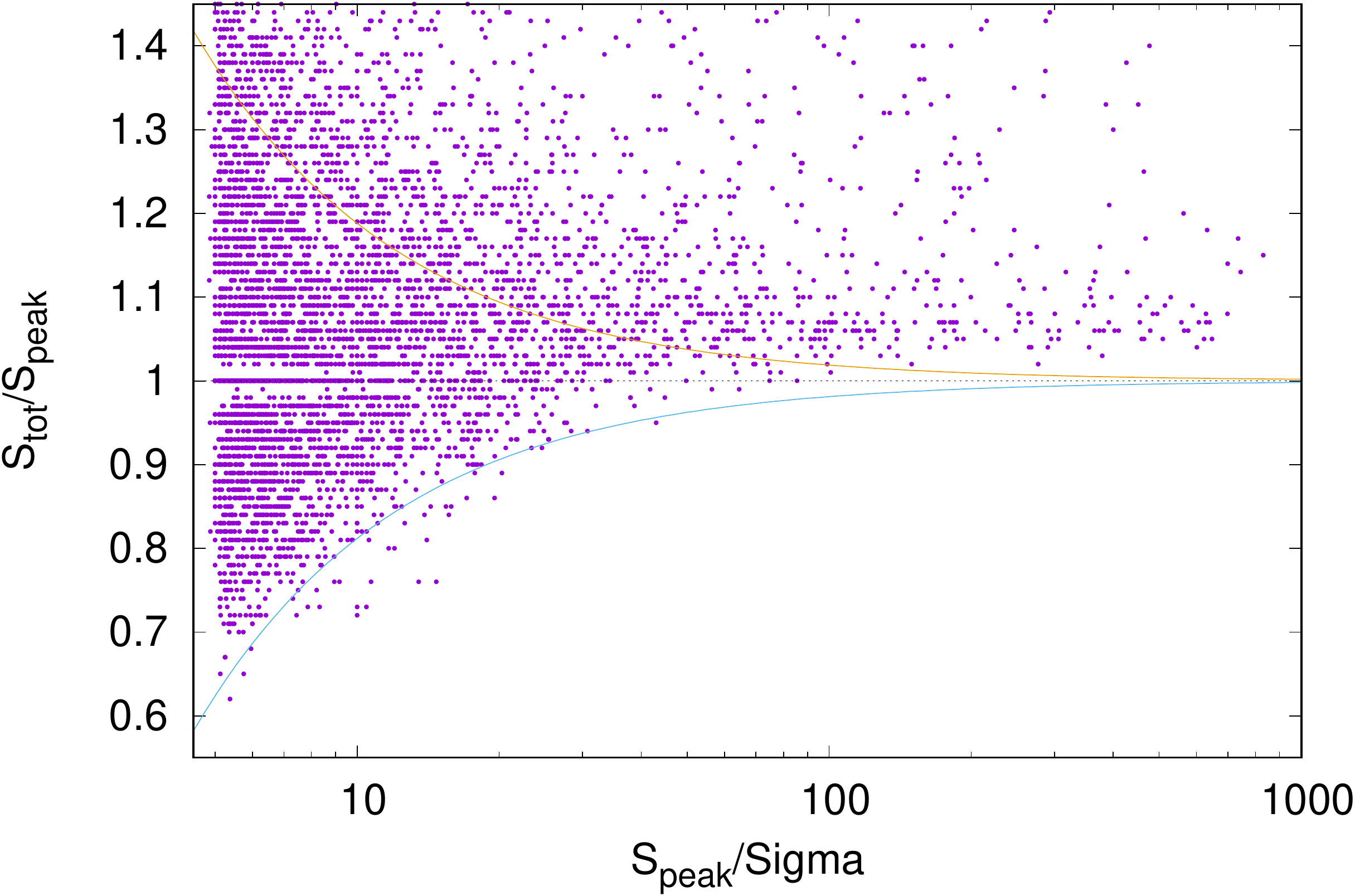}}
\caption{The plot showing the ratio of the integrated flux density to peak intensity as a function of the significance of detection. Inorder to highlight the distribution near the ratio of 1, the y-axis has been restricted in the range 0.5 to 1.5.}
\label{fig:en1wq}
\end{figure}

In Figure~\ref{fig:en1wq} we plot the ratio of total flux density to peak intensity against the
significance of detection ($S_{peak}/\sigma$). In order to fit an
envelope to $S_{peak}/\sigma$ as a function of source flux, we divided
the data into several bins with closer spacing at lower significance and
fewer bins at higher significance to check the distribution of sources.
In each bin there were a few out-liers with $S_{\rm tot}/S_{\rm peak} <
1$. To exclude these out-liers, we considered the ratio of total to peak
where the distribution started to become continuous and this value was
taken as the lower envelope. We fitted a profile to this envelope with
the function $1.0 + (1.88 \pm 0.10)/x$, where $x$ is $S_{peak}/\sigma$.
The envelope is shown in the figure. All the points that fall within this
envelope on both sides are  taken as point sources. With this criterion,
about 60 per cent of sources are unresolved.  If we were to consider the
total to peak flux ratio $<1.5$ as compact sources then the number of
compact sources are 4,790  ($\sim 75$ per cent).

Table~\ref{tab:catalogue} lists the ten entries of the source
catalogue.  The columns contain:
(1) the source name JHHMMSS+DDMMSS, where HHMMSS are the hours, minutes
and seconds of time of right ascension and DDMMSS are the degrees,
minutes and seconds of arc of declination (both in J2000),
(2) the J2000 right ascension (degrees) and error (arcseconds),
(3) the J2000 declination (degrees) and error (arcseconds),
(4) the integrated Stokes $I$ flux density and error (mJy),
(5) the peak Stokes $I$ intensity and error (mJy beam$^{-1}$),
(6) the rms at the source position (mJy) returned by PyBDSF,
(7 and 8) the dimensions of the fitted major and minor axes of 2D
Gaussian with errors (arc seconds),
(9) the position angle of the source major axis (degrees) with respect
to north,
and (10) the source classification code, as in \citep{PyBDSF2015}.
All errors are $1\sigma$.  Errors on flux density include a contribution
of 2.6\% derived as the percent rms variation of the calibration 
solution for the flux of the secondary calibrator 1549+506 over the
23 different observing sessions.  This error was added in quadrature to
the PyBDSF measurement error. The full source catalogue is available in machine readable form.

\subsection{Spectral index of compact sources}

\begin{figure}
\centerline{\includegraphics[width=3.5in,angle=0]{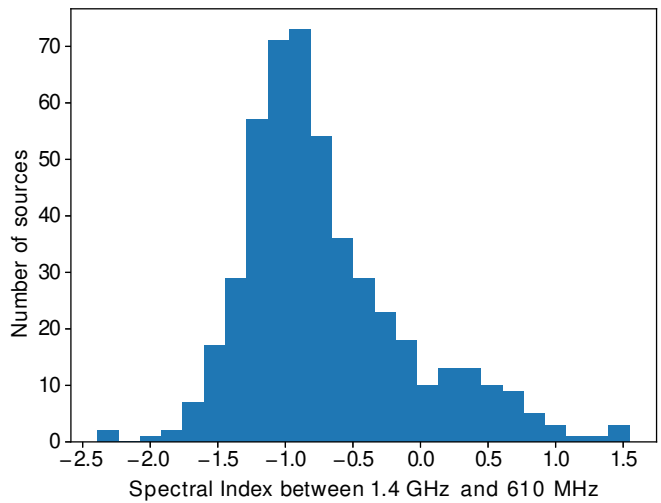}}
\caption{The spectral index distribution between 610 MHz and 1.4 GHz.
The median spectral index is $-0.85 \pm 0.05$.}
\label{fig:en1wspix}
\end{figure}

We have used the compact sources ($S_{\rm tot}/S_{\rm peak} < 1.5$) to
check for spectral index {w.r.t.} VLA FIRST at 1.4 GHz and the GMRT 325
MHz catalogue from \citep{2009MNRAS.395..269S}. The median
spectral index {w.r.t.} VLA FIRST 1.4 GHz is $-0.85 \pm 0.05$ and {w.r.t.}  325 MHz  is $-0.73 \pm 0.12$. The spectral index
distribution between our 610 MHz and the VLA FIRST 1.4 GHz observations
is shown in Figure~\ref{fig:en1wspix}. The median 610 MHz flux density of sources with a VLA FIRST counterpart is 4.5 mJy, whereas the median 610 MHz flux
density of sources with a GMRT 325 MHz counterpart is 0.72 mJy. 
The GMRT 325 MHz observations  covers a smaller region ($\sim 1.5$ deg$^2$). 
 Recently, higher sensitivity data from the upgraded GMRT at band-3 (250-500 MHz)
has been published \citep{2019MNRAS.490..243C}. The median spectral index using these data is $0.70 \pm 0.18$. 
The steeper spectral index for brighter sources and relatively flatter
spectral index for fainter sources is also seen in other samples
\citep{2013MNRAS.429.2080W, 2016MNRAS.463.2997M}. 

\subsection{Source Counts}

The large area and sub-mJy depth of the observations allows us to derive
source counts with low statistical errors down to $\sim 200$ $\mu$Jy.
However systematic effects become a significant factor at intermediate
signal-to-noise levels.  For flux limited observations in the presence
of noise, the two important effects are Eddington bias
\citep{1913MNRAS..73..359E}, and  incompleteness of the sample at faint
flux densities.  Eddington bias arises due to
the effects of the noise distribution on the measured flux densities of
the underlying ensemble of radio sources, which distorts the observed
flux density distribution relative to the true flux density distribution.
Eddington bias corrections generally assume that the noise is well
described by a normal distribution \citep{1940MNRAS.100..354E}. However
for wide-field radio mosaic images the effect of image noise on the
detected source fluxes will differ from a simple normal distribution,
due both to the change in background image rms with position arising
from the spatial variation  of the mosaic weights, and to the
non-Gaussian increase of the background rms in the vicinity of bright
sources caused by residual deconvolution errors.

The observed differential source counts can be related to the true
source counts as
\begin{equation}
  \frac{{\rm d}N_{\rm o}(s')}{{\rm d}s'} =
  \int_0^\infty \frac{{\rm d}N_{\rm t}(s)}{{\rm d}s} \, \epsilon(s) \,
  p(s,s') \, {\rm d}s
  \label{eqn:counts}
\end{equation}
Here ${\rm d}N_{\rm o}(s')/{\rm d}s'$ is the observed count at the observed
flux densities $s'$, and ${\rm d}N_{\rm t}(s)/{\rm d}s$ is the true source
count  at the true flux density $s$. The function $p(s,s')$ is the normalized
probability density function that a source at observed flux $s'$ is due
to a source with true flux density $s$, and $\epsilon(s)$ is the
probability that a source with true flux density, $s$, will result in a
detection.

\begin{figure}
\centerline{\includegraphics[width=3.6in]{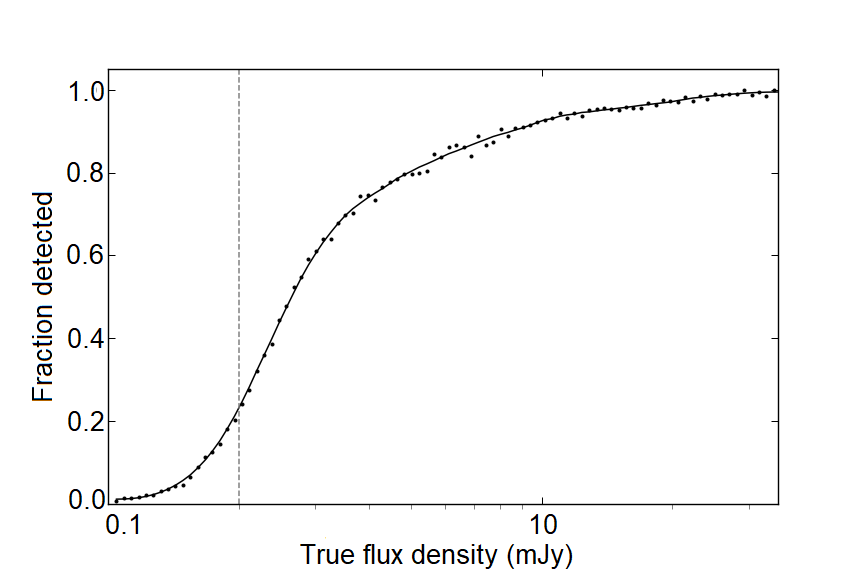}}
\caption{Fraction of sources detected as a function of
true flux density. The data points are the results of simulations
described in the text. The solid line is model fit based on
piece-wise polynomial spline interpolation of the simulated data.
The vertical dashed line shows the approximate $5\sigma$ detection level of 200 $\mu$Jy beam$^{-1}$.}
\label{fig:efficiency}
\end{figure}

We measure both $p(s,s')$ and $\epsilon(s)$ by injecting sources into
the PyBDSF residual image - the image with the fits to the detected
radio sources removed.  For true flux density $s$ ranging from 
below the noise to several mJy, we injected 3000 sources at
fixed values of true flux density with the dimension of
the synthesized beam.
We do not consider resolved sources in this analysis as we
expect the dominant source population at sub-mJy flux densities
is a combination of star forming galaxies and radio quiet AGN 
\citep{Ocran2020}, which
will be unresolved at angular resolution of several arc seconds.
These injected sources populate an image with the same
background noise and rms properties as the original source finding. We
then ran PyBDSF on the images with injected sources and measured the
fraction of sources detected as a function of true flux, and the
distribution of true flux densities of injected sources that result in a
source with observed flux $s'$. Figure~\ref{fig:efficiency} shows the
detection efficiency function $\epsilon(s)$.  The data points are the
results for each simulation of 3000 sources, and the solid line is a
piece-wise polynomial spline interpolation of the data. The vertical
dashed line in the Figure \ref{fig:efficiency} is the approximate $5\sigma$ detection
threshold. Sources with true flux well below the threshold can produce
detections because some fraction of the sources have their `observed'
flux density increased above the threshold according to $p(s,s')$.
As expected the functions $p(s,s')$ are generally non-Gaussian in shape.

\begin{figure}
\centerline{\includegraphics[width=3.6in]{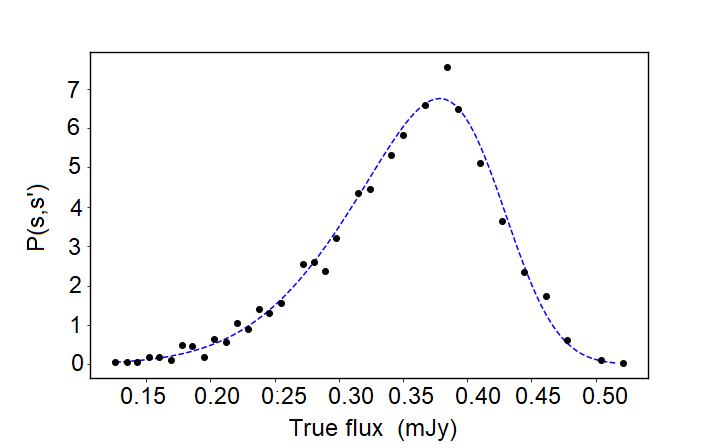}}
\caption{An example probability density distribution $p(s,s')$ for $s'
= 380$ $\mu$Jy.
The dashed line is a skew normal distribution fit to the simulations
(data points).}
\label{fig:true_pdf}
\end{figure}

\begin{figure*}
\centerline{\includegraphics[width=6.5in,angle=0]{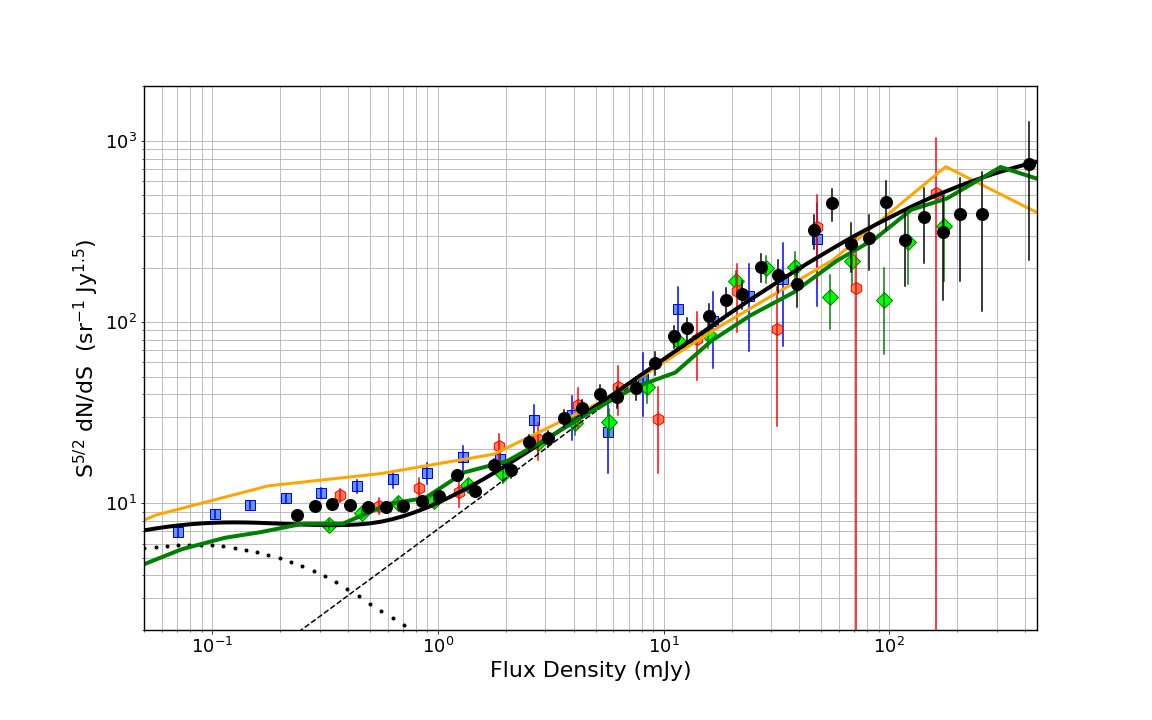}}
\caption{Radio source counts at 610 MHz. Counts from this work are shown as filled black circles. Red hexagons are counts from
\protect\cite{2007A&A...463..519B}, and green diamonds are from \protect\cite{2008MNRAS38375G}. The blue squares are the recent results of very deep GMRT imaging by \protect\cite{Ocran2020}.
For comparison, the solid lines show model predictions for 610\,MHz counts from \protect\cite{Wilman2008} (green), \protect\cite{2010MNRAS.404..532M} (black), and \protect\cite{Bonaldi2019} (orange). The dotted and short dashed lines show indicative model counts for starburst galaxies
and AGN from \protect\cite{2010MNRAS.404..532M}.}
\label{fig:counts}
\end{figure*}

Figure~\ref{fig:true_pdf} shows an example of the shape of the function
for the case of $s' = 380$ $\mu$Jy beam$^{-1}$. The function is skewed,
with a negative tail toward lower flux densities. We fit a skew normal
distribution to the data as shown by the dashed line in
Figure~\ref{fig:true_pdf}.  A skew normal is described as
\begin{equation}
    f(x) = \frac{2}{\omega}\phi(x)\Phi(\gamma x)
\end{equation}
where
\begin{equation}
  \phi(x) = \frac{1}{\sqrt{2\pi}}{\rm e}^{-\frac{x^2}{2}},
  \quad {\rm and} \quad
  \Phi(\gamma x) = \frac{1}{2} \left[ 1 +
   {\rm erf} \left( \frac{\gamma x}{\sqrt{2}} \right) \right]
\end{equation}
The skew parameter, $\gamma$, is zero for a symmetric normal
distribution. The parameters of the skew normal distribution are fit to the data as a function of $s'$, so $\gamma = \gamma(s')$ and
\begin{equation}
  x = \frac{s - \xi(s')}{\omega(s')}
\end{equation}
This allows the probability density distribution $p(s,s')$ to be
constructed for any value of $s'$ and $s$, and together with
$\epsilon(s)$ allows an predicted observed source count to be derived
from the a true source count using the integral
Equation~\ref{eqn:counts}. We solve for the combined completeness and
bias correction by first setting the true count equal to the observed
count and iteratively modifying the trial true count until the left hand
side of Equation~\ref{eqn:counts} equals the observed counts. We then
derive a per-bin correction to the observed counts. The resulting source
counts are listed in Table~\ref{tab:src_cnt} and plotted in
Figure~\ref{fig:counts}. 
Our counts lie near the middle of the scatter of other observed counts at 610 MHz \citep{2007A&A...463..519B,2008MNRAS38375G,Ocran2020}.
The scatter of the observations of about a factor of two below 1\,mJy,
is similar to the scatter in
model counts shown on the plot \citep{Wilman2008,2010MNRAS.404..532M,Bonaldi2019}.

\begin{table}
\centering
\caption{Euclidean normalized 610 MHz source counts within the  ELAIS\,N1 Wide field}
\begin{tabular}{|c|c|c|r|r|}
   \hline
 Mean Flux & Bin Width & $N$ & Count\,& Error   \\
 (mJy) & (mJy) & & (Jy$^{1.5}$\,sr$^{-1}$)&(Jy$^{1.5}$\,sr$^{-1}$)\\
   \hline
   0.239 &  0.043   & 510 &   7.65 &   0.34 \\
   0.285 &  0.052   & 664 &   8.55 &   0.33 \\
   0.341 &  0.062   & 705 &   8.80 &   0.33 \\
   0.408 &  0.075   & 626 &   8.63 &   0.35 \\
   0.489 &  0.090   & 525 &   8.46 &   0.37 \\
   0.589 &  0.107   & 412 &   8.44 &   0.42 \\
   0.702 &  0.129   & 344 &   8.58 &   0.46 \\
   0.849 &  0.155   & 289 &   9.14 &   0.54 \\
   1.010 &  0.186   & 246 &   9.77 &   0.62 \\
   1.223 &  0.223   & 237 &  12.64 &   0.82 \\
   1.459 &  0.267   & 151 &  10.44 &   0.85 \\
   1.766 &  0.321   & 155 &  14.39 &   1.16 \\
   2.113 &  0.385   & 112 &  13.57 &   1.28 \\
   2.538 &  0.462   & 121 &  19.30 &   1.75 \\
   3.063 &  0.555   & 95  &  20.22 &   2.07 \\
   3.641 &  0.666   & 96  &  26.22 &   2.68 \\
   4.370 &  0.799   & 83  &  29.82 &   3.27 \\
   5.208 &  0.958   & 77  &  35.75 &   4.07 \\
   6.213 &  1.150   & 57  &  34.27 &   4.54 \\
   7.554 &  1.380   & 47  &  38.39 &   5.60 \\
   9.147 &  1.656   & 48  &  52.70 &   7.61 \\
  11.113 &  1.987   & 50  &  74.44 &  10.53 \\
  12.743 &  2.385   & 47  &  82.10 &  11.99 \\
  15.893 &  2.862   & 38  &  96.10 &  15.59 \\
  18.897 &  3.434   & 36  & 116.92 &  19.49 \\
  22.314 &  4.121   & 31  & 127.16 &  22.84 \\
  27.144 &  4.945   & 32  & 178.52 &  31.56 \\
  32.047 &  5.934   & 23  & 161.94 &  33.77 \\
  39.030 &  7.121   & 15  & 144.07 &  37.20 \\
  46.524 &  8.546   & 23  & 285.57 &  59.55 \\
  55.476 & 10.255   & 25  & 401.63 &  80.33 \\
  67.533 & 12.306   & 11  & 240.79 &  72.60 \\
  81.071 & 14.767   &  9  & 259.22 &  88.41 \\
  96.641 & 17.720   & 11  & 409.62 & 123.50 \\
  \hline
\end{tabular}
\label{tab:src_cnt}
\end{table}

\subsection{Multi-Wavelength counterparts and redshift distribution}

\begin{figure}
\centerline{\includegraphics[width=3.0in,angle=0]{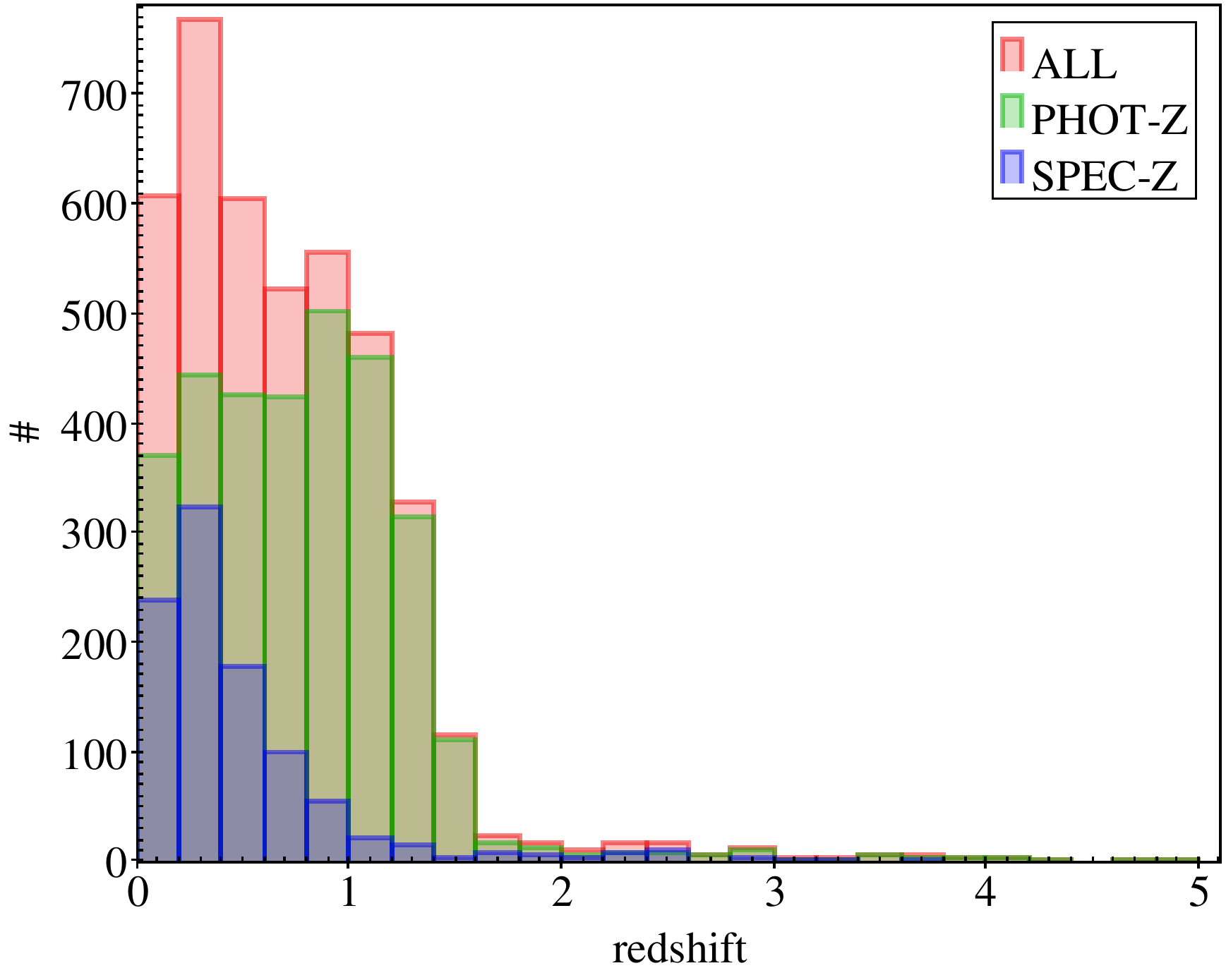}}
\caption{The Redshift Distribution of GMRT Sources.}
\label{fig:redshift}
\end{figure}

The extensive multi-wavelength data available within the ELAIS\,N1 field has been collected and homogeneized over the last few years. The Spitzer Data Fusion
\citep{Vaccari2010,Vaccari2015}\footnote{\url{http://mattiavaccari.net/df}} project has produced Spitzer-selected multi-wavelength catalogues for 8 extragalactic survey fields and a total of 65 deg$^2$ covered in the 7 IRAC and MIPS bands. The Herschel Extragalactic Legacy Project
\citep{Vaccari2016,2018A&A...620A..50M,2019MNRAS.490..634S}\footnote{\url{https://herschel.sussex.ac.uk}} has produced multi-wavelength catalogues for sources over the 1,300 deg$^2$ of the extragalactic sky covered by Herschel. 
 For this work, we have used data products from both projects to optimize the multi-wavelength source characterization process. In order to minimize the ambiguity of the identification, we have matched radio sources with their Spitzer/SWIRE mid-infrared and UKIDSS DXS near-infrared (where a Spitzer/SWIRE counterpart was not found) counterparts using a nearest-neighbour technique with a search radius equal to three times the estimated astrometric error. 
We have combined the spectroscopic redshift compilations from the Spitzer Data
Fusion\footnote{\url{http://mattiavaccari.net/df/specz}} and
HELP\footnote{\url{http://hedam.lam.fr/HELP/dataproducts/dmu23/}} projects to collect spectroscopic redshifts. Where a spectroscopic redshift was not available, we have used photometric redshifts from (in order of preference) the HSC \citep{Tanaka2018}, SWIRE \citep{RowanRobinson2008,RowanRobinson2013}, HELP \citep{Duncan2018a,Duncan2018b},
SERVS \citep{Pforr2019} and SDSS \citep{Beck2016} projects. The redshift distribution is illustrated in Figure~\ref{fig:redshift}.

Multi-wavelength coverage and matching statistics are summarized in Table~\ref{tab:matches}. We find that 86.3\% of the GMRT footprint is covered by either SWIRE or UKIDSS, and within this common footprint we find that 91.9\% of our sample finds a match and 74.4\% a redshift estimate. At the flux density limits of our radio survey and of our multi-wavelength datasets, we expect the overwhelming majority of radio sources to have an infrared counterpart, and indeed the vast majority of unmatched sources fall outside the SWIRE and UKIDSS coverage. Of the $\sim 8$\% sources which does not have counterparts, the majority of them could be the Infrared Faint Radio Sources (IFRS) \citep{2011ApJ...736...55N}, however, some of them could also be the lobes of active galaxies. We have not investigated unmatched radio sources within
the SWIRE/UKIDSS coverage individually as a full multi-wavelength analysis is outside the scope of the present paper. However, a cursory inspection of a few unmatched radio sources confirms that, as also found
by \cite{Ocran2017,Ocran2020}, most of them are  in `complex' areas of the radio or of the infrared images and thus do not lend themselves to being identified via an automated procedure.

\begin{table}
\centering
\caption{Multi-Wavelength Data for GMRT Sources.}
\begin{threeparttable}
\begin{tabular}{lrr}\hline
  Catalogue                 & No. of Sources & Fraction \\
  GMRT            &      6400 &   100$\%$ \\
  COVERED-ALL     & 5523/6400 &  86.3$\%$ \\
  COVERED-SWIRE   & 5473/6400 &  85.5$\%$ \\
  COVERED-UKIDSS  & 5024/6400 &  78.5$\%$ \\
  MATCHED-ALL     & 5078/5523 &  91.9$\%$ \\
  MATCHED-SWIRE   & 4945/5473 &  90.4$\%$ \\
  MATCHED-UKIDSS  & 4029/5024 &  80.2$\%$ \\
  SWIRE IRAC1234  & 2827/5523 &  51.2$\%$ \\
  SWIRE MIPS24    & 3408/5523 &  61.7$\%$ \\
  X-RAY           &   59/5523 &   1.1$\%$ \\
  SPECZ           &  970/5523 &  17.6$\%$ \\
  PHOTZ           & 3138/5523 &  56.7$\%$ \\
  REDSHIFT (ANY)  & 4108/5523 &  74.4$\%$ \\ \hline
\end{tabular}
\end{threeparttable}
\label{tab:matches}
\end{table}

\begin{figure*}
\begin{center}
\begin{tabular}{llll}
 \includegraphics[width=1.5in]{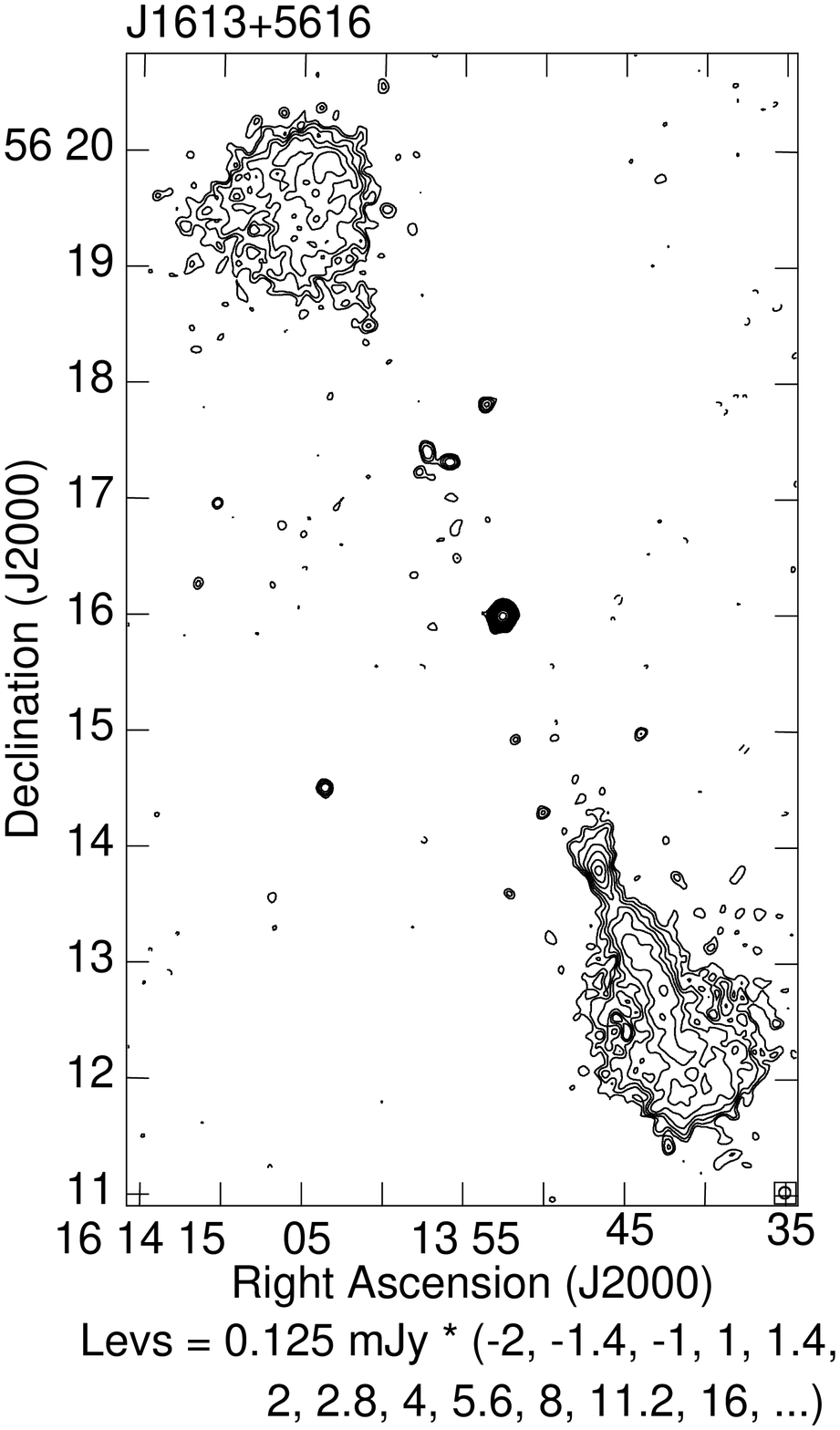} &
 \includegraphics[width=1.6in]{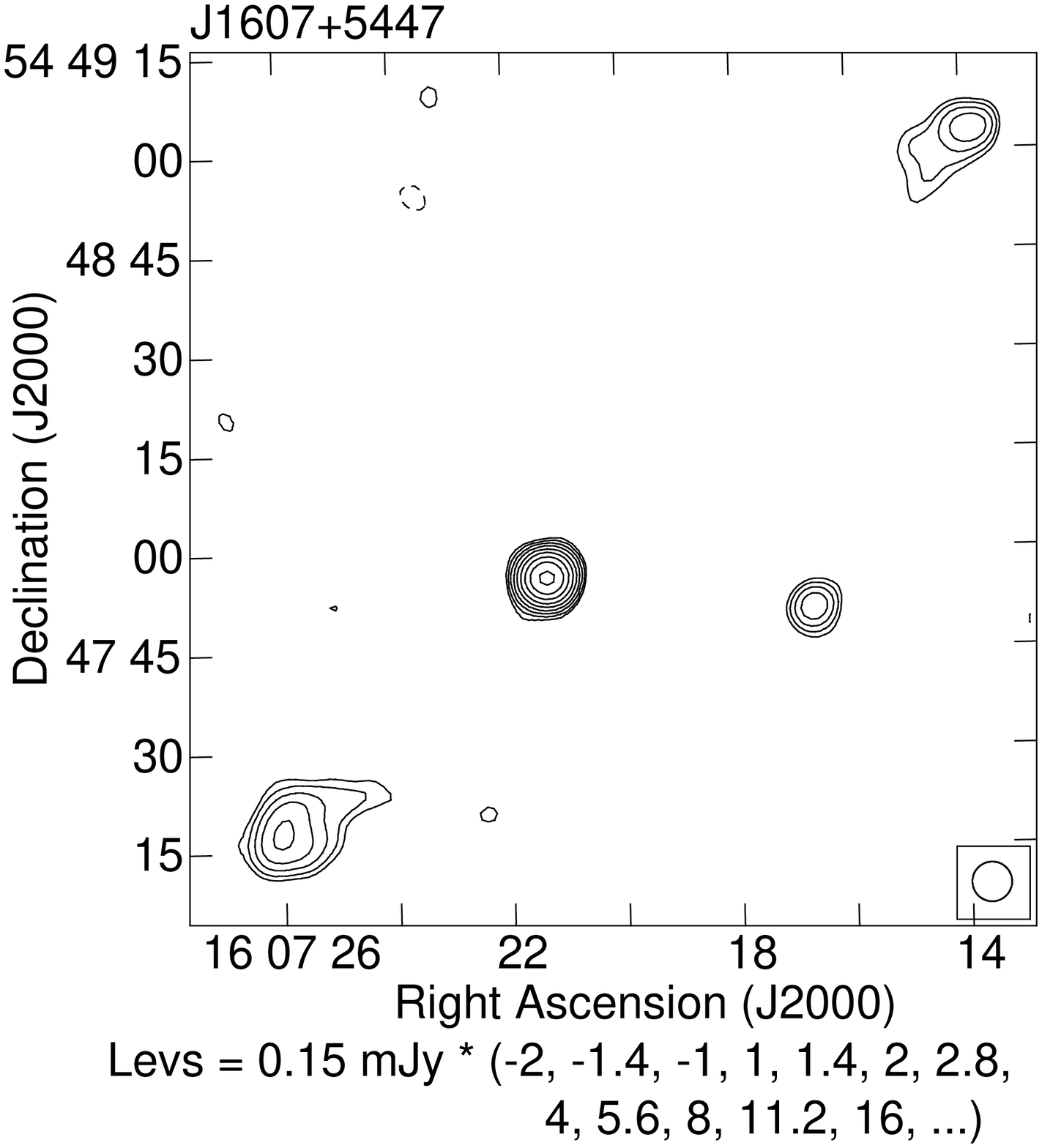}  &
 \includegraphics[width=1.6in]{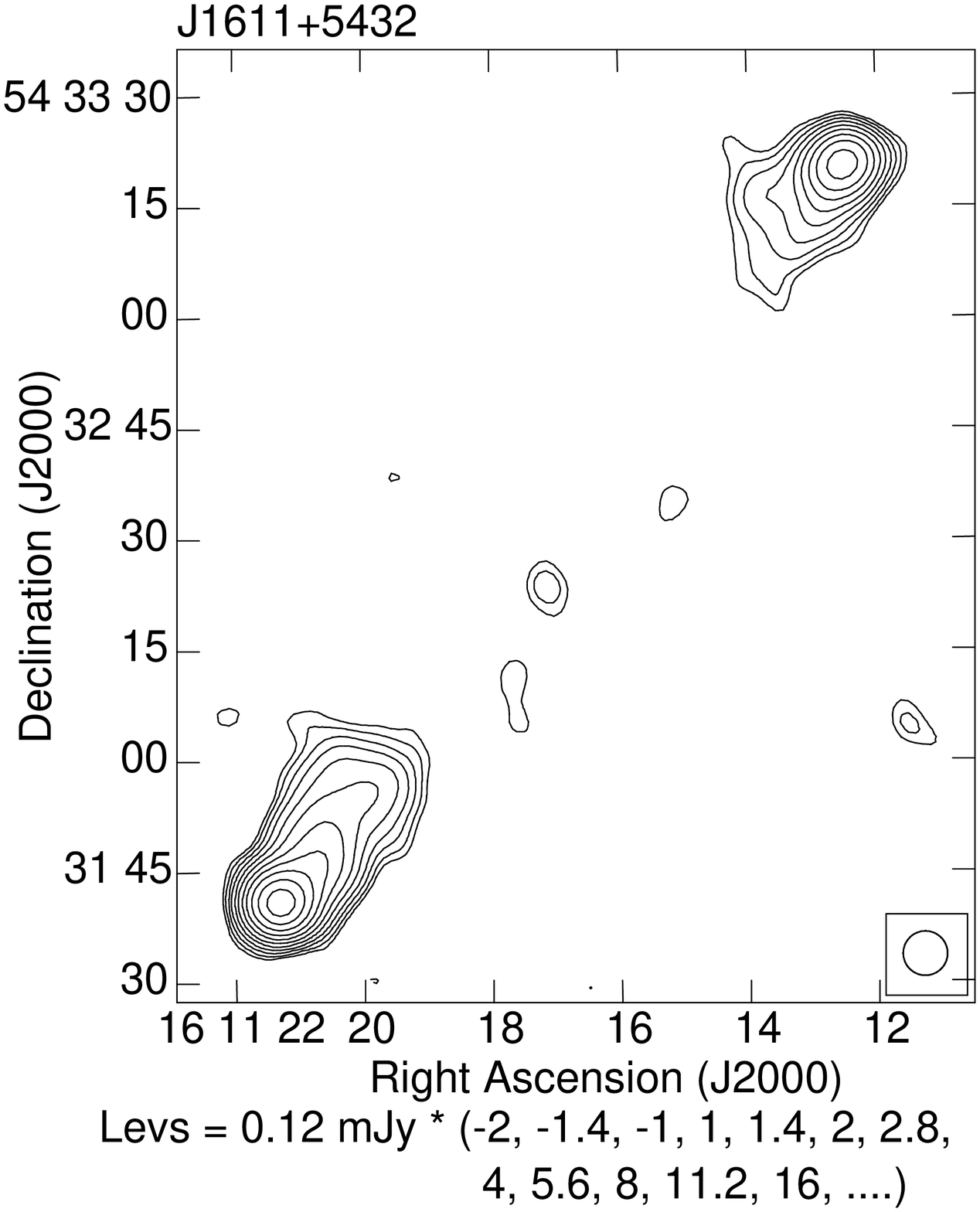} &
 \includegraphics[width=1.5in]{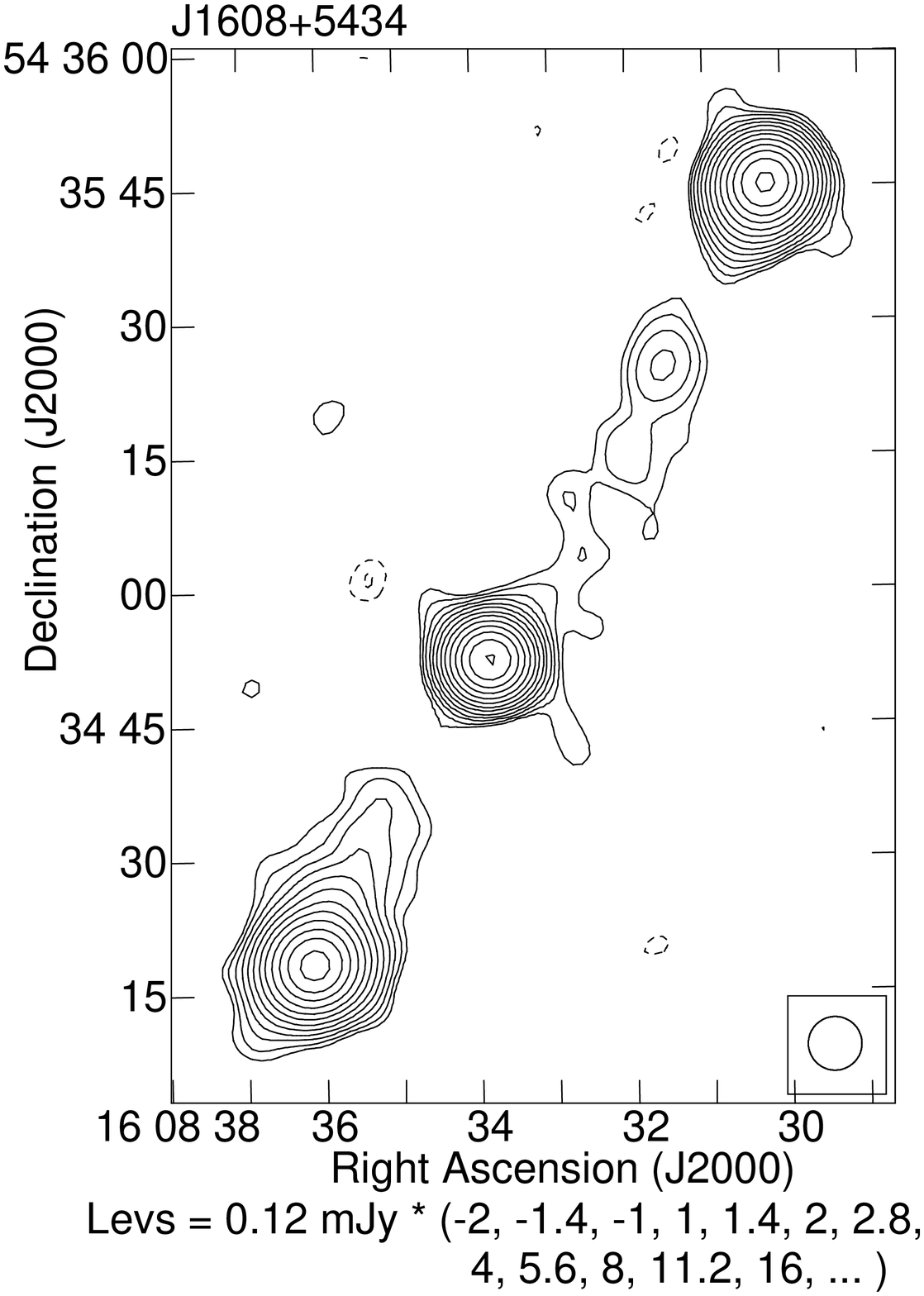}
\end{tabular}
\vspace{0.25in}
\begin{tabular}{lll}
 \includegraphics[width=2.10in]{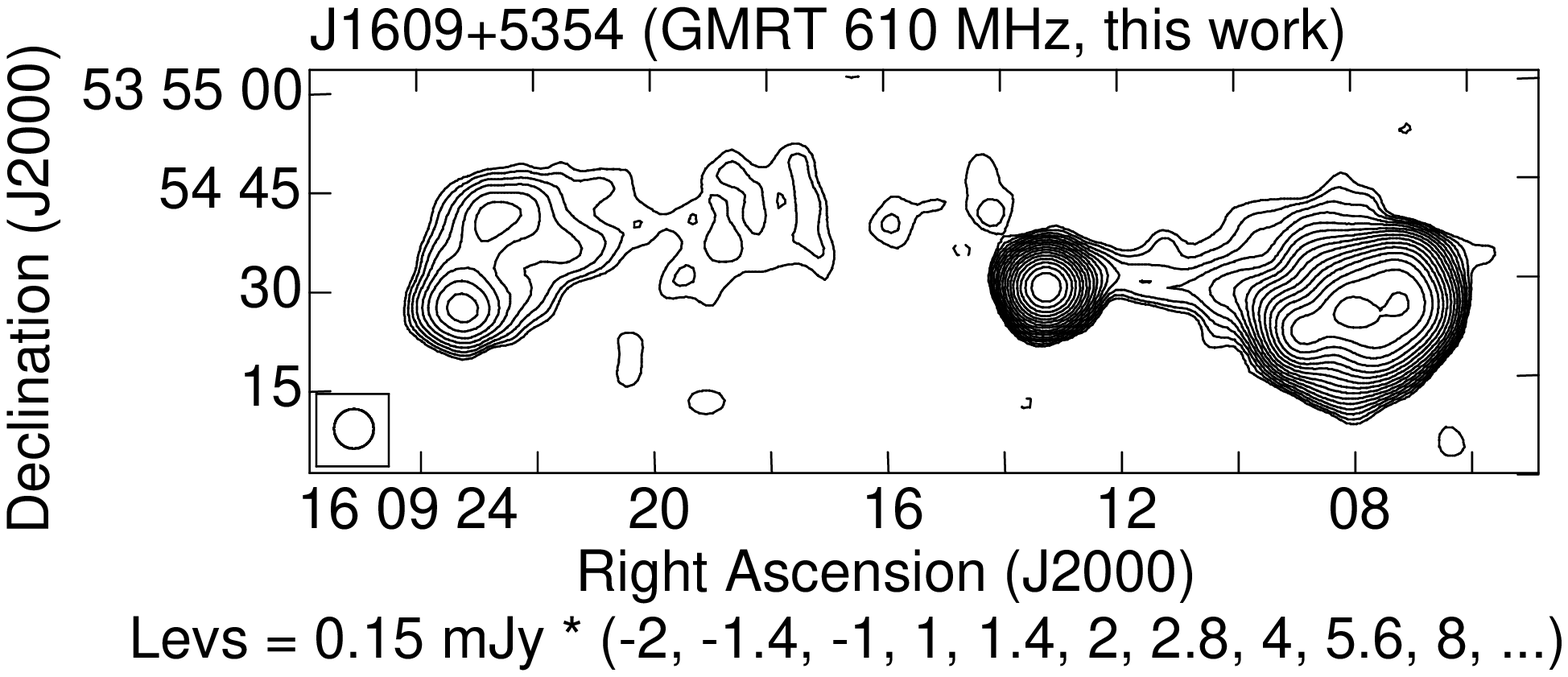}  &
 \vspace{0.25in}
 \includegraphics[width=2.10in]{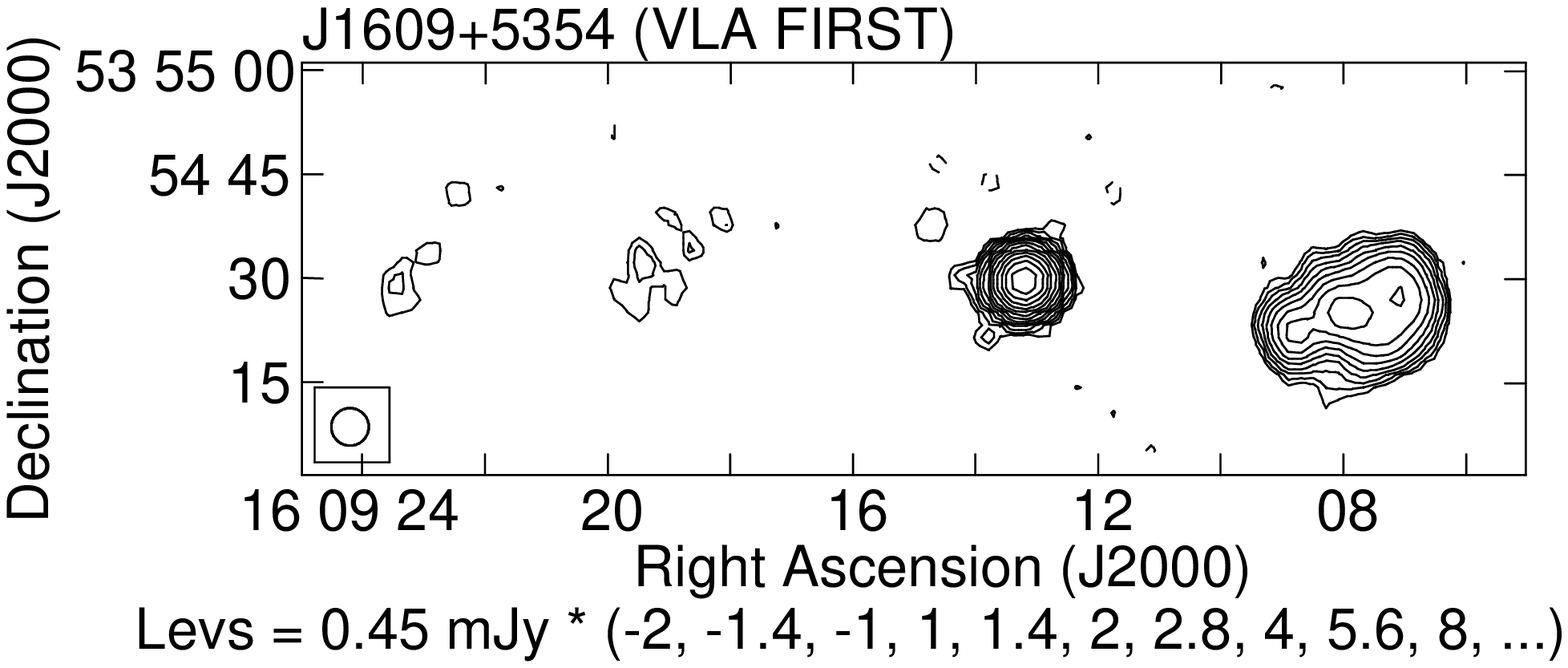} &
 \vspace{0.25in}
 \includegraphics[width=1.9in]{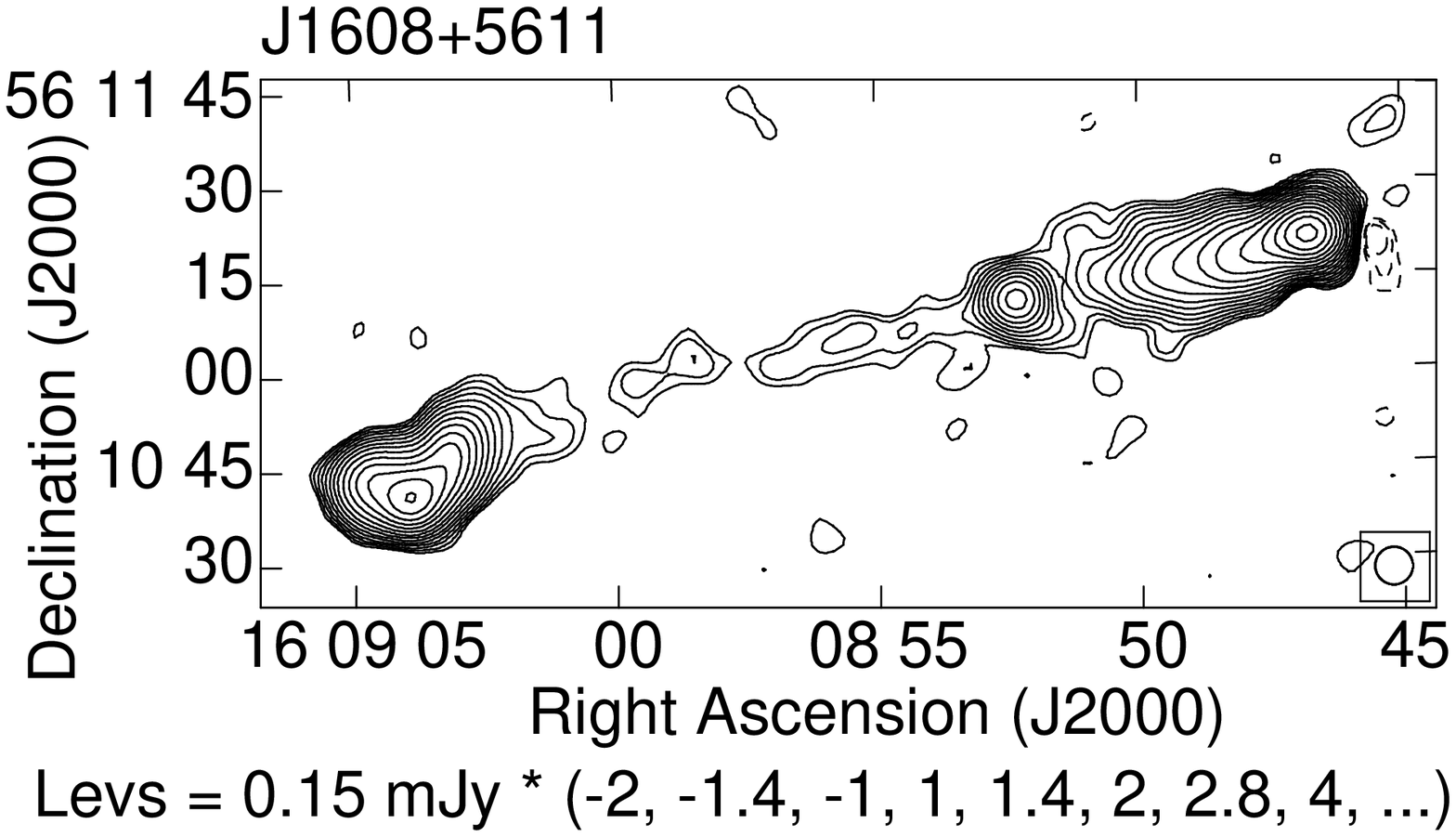} \\
\end{tabular}
\end{center}
\vspace{-0.5in}
\caption{Giant Radio Sources discovered in our ELAIS\,N1 wide area
survey. Top row from left: Giant radio galaxies J1613+5616, 1607+5447, J1611+5432 and J1608+5434. Bottom row: Giant radio quasars J1609+5354 (GMRT 610 MHz), J1609+5354 (VLA FIRST 1.4 GHz) and J1608+5611.}
\label{fig:GRG1}
\end{figure*}
\section{Giant Radio Sources}

 During the visual inspection of the ELAIS\,N1 field for large angular sized sources, we have discovered several giant radio sources (GRS) with linear size of over a Mpc. 
GRS are believed to be the late stage in the evolution of radio galaxies \citep{1999MNRAS.309..100I} and hence finding more number of GRSs in the sky is important to shed light on possible scenarios that can explain
the formation and evolution of GRS \citep{2020A&A...635A...5D}. The number of known GRS have dramatically increased in recent years due to careful investigation of wide area sky surveys in the radio and availability of redshifts from optical surveys such as  SDSS \citep{2016ApJS..224...18P,2017MNRAS.469.2886D,2020A&A...635A...5D}. The apparent low space density of GRSs may be due to lack of sensitive wide area surveys \citep{1997MNRAS.292..723K}.

In the ELAIS\,N1 field, we found five objects with linear size of  $\sim$ a Mpc or larger and one object with a linear size of 0.85 Mpc. This is among the highest number density of GRSs in a deep field, comparable to the number density of GRSs in \citet{2020A&A...635A...5D}. Most of the GRS discovered from this work are too faint to be detected in the existing surveys such as VLA FIRST and NVSS. It appears now that GRSs are more common than indicated by earlier studies \citep{1997MNRAS.292..723K}, highlighting the importance of sensitive wide area survey to discover more such objects.

Here we briefly describe the properties of individual giant sources. The redshifts were obtained from NASA/IPAC Extragalactic Database (NED) at the position of the core.

\subsubsection*{J1613$+$5616}

This is a low redshift GRS at $z=0.18$ (Figure \ref{fig:GRG1}, top left panel). The angular size is $\sim 8.5$ arcmin  which makes the linear size 1.5 Mpc. 
The flux density of the core is 11.5 mJy at 610 MHz, 7.8 mJy from VLA
FIRST at 1.4 GHz and 8.7 mJy in NVSS at 1.4 GHz. 
Due to the large angular size of the object and high angular resolution, the lobes are highly resolved, and hence the total flux density may be underestimated due to missing flux. The spectral index of the core is
flat suggesting a currently active AGN, despite its large size. Assuming a 
hotspot advance speed of $\sim 0.1c$ gives the kinematic age as $\sim
50$ million years. Midway to the hotspots, two peaks are seen roughly
equidistant from the core, at $16^{\rm h}13^{\rm m}57\fs3$ and
$+56^{\circ}17'23\farcs6$ in the northern side and at $16^{\rm h}13^{\rm
m}46\fs6$ and $+56^{\circ}13'47\farcs8$ in the southern direction. Such
structures could normally suggest the source to be a double-double radio
source, however it is puzzling to see this `inner pair' much weaker than
the outer pair, if this is true. The absence of bright outer hotspots,
flat-spectrum core and large kinematic age does support the
double-double source possibility. Spectral index information of the inner and outer structures are needed to conclude the double-double nature of the source.

\subsubsection*{J1607$+$5447}

The angular size of the source is 2.5$'$  (Figure \ref{fig:GRG1}, top panel, second from left). For this  size, it will be  a GRS with size $>1$ Mpc for any redshift above 0.6. The HSC Photometric
redshift of this source is 1.14 and the corresponding linear size of the
source is 1.25 Mpc,  classifying this as a giant radio source. The lobes are faint and the emission is confined to largely near the hotspots. In the VLA FIRST survey, only the core is detected, with a flux density of 2.5 mJy.
The flux density of the core at 610 MHz is 2.8 mJy.

\begin{figure*}
\begin{center}
\begin{tabular}{lll}
\includegraphics[width=2.0in]{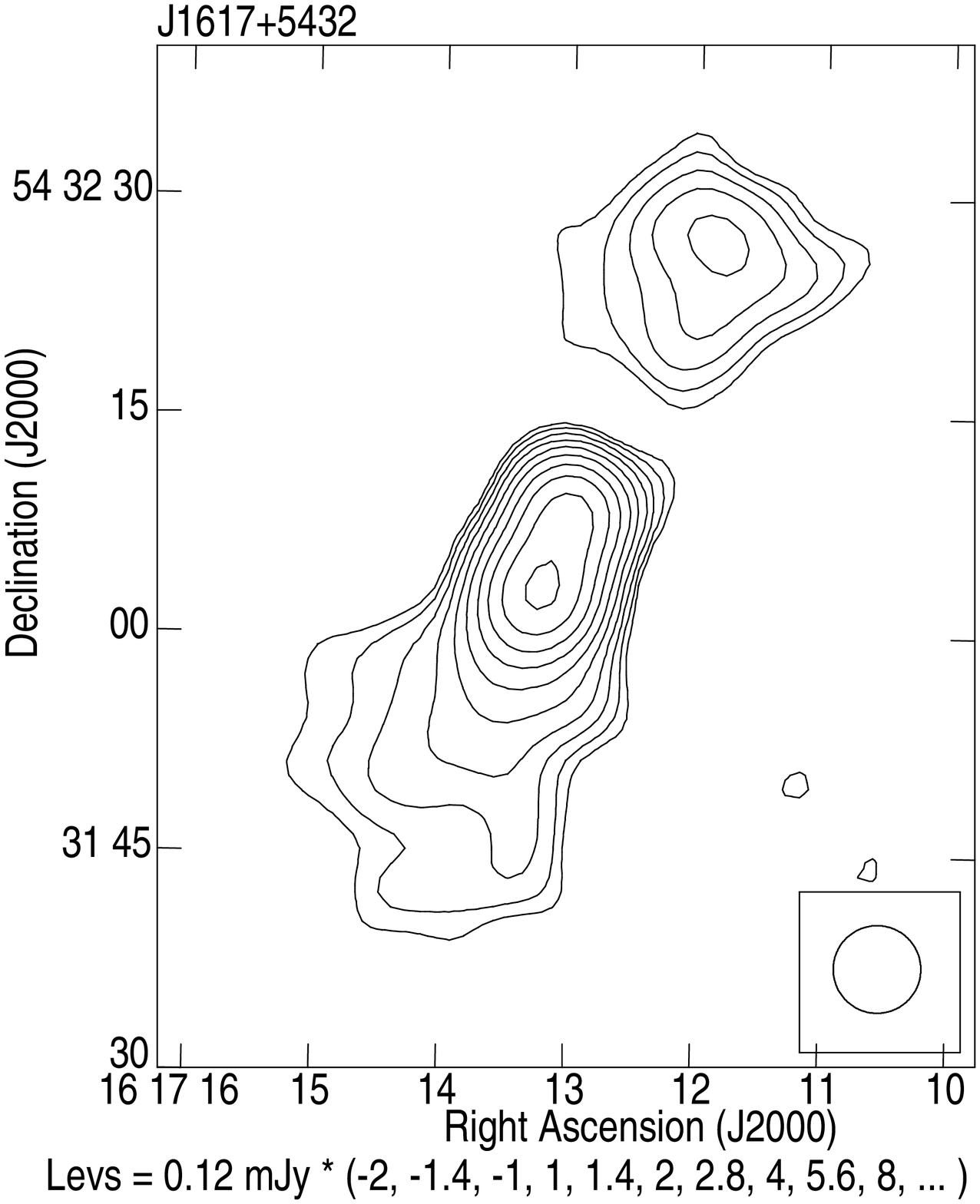} &
\includegraphics[width=2.3in]{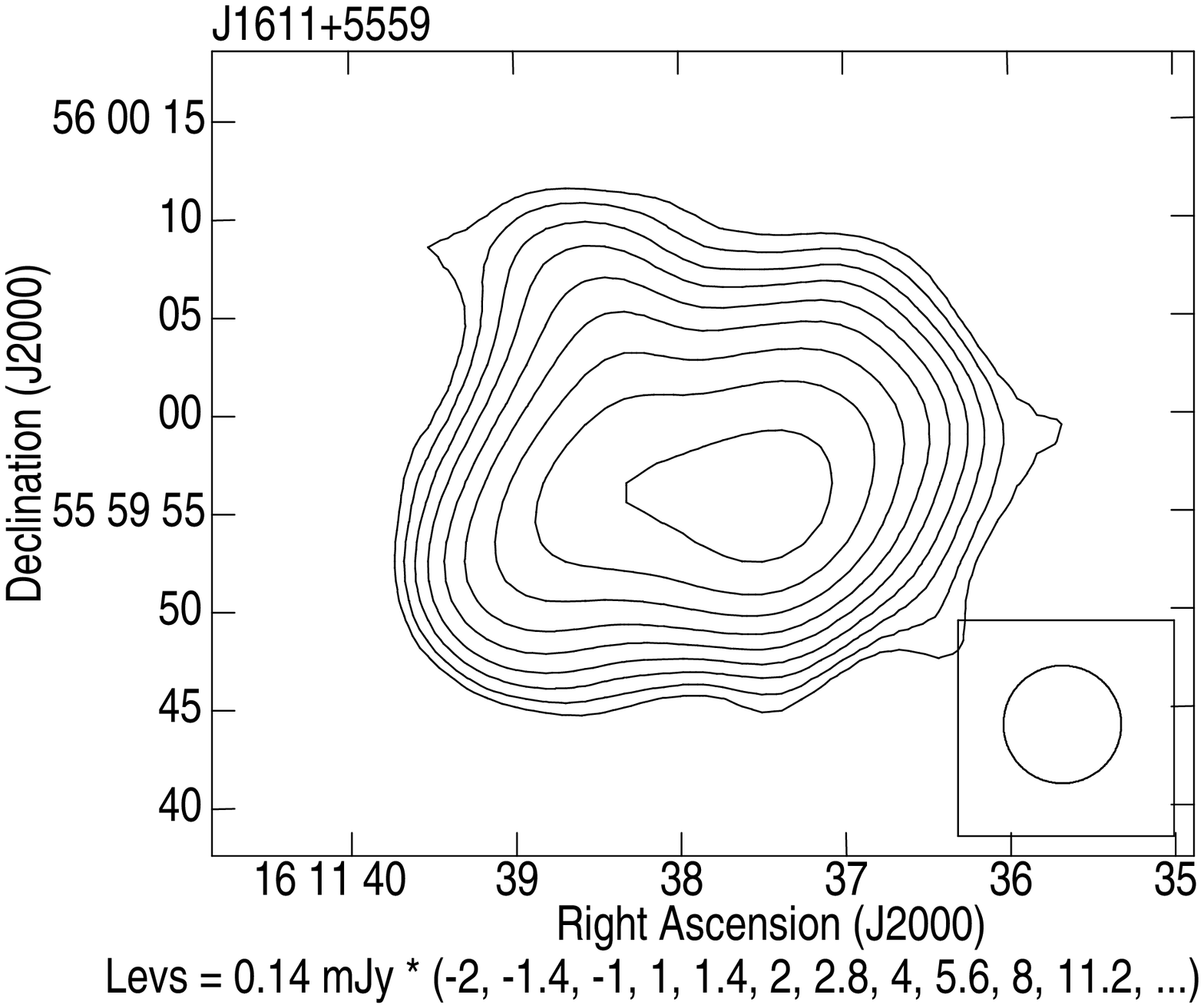}  &
\includegraphics[width=2.0in]{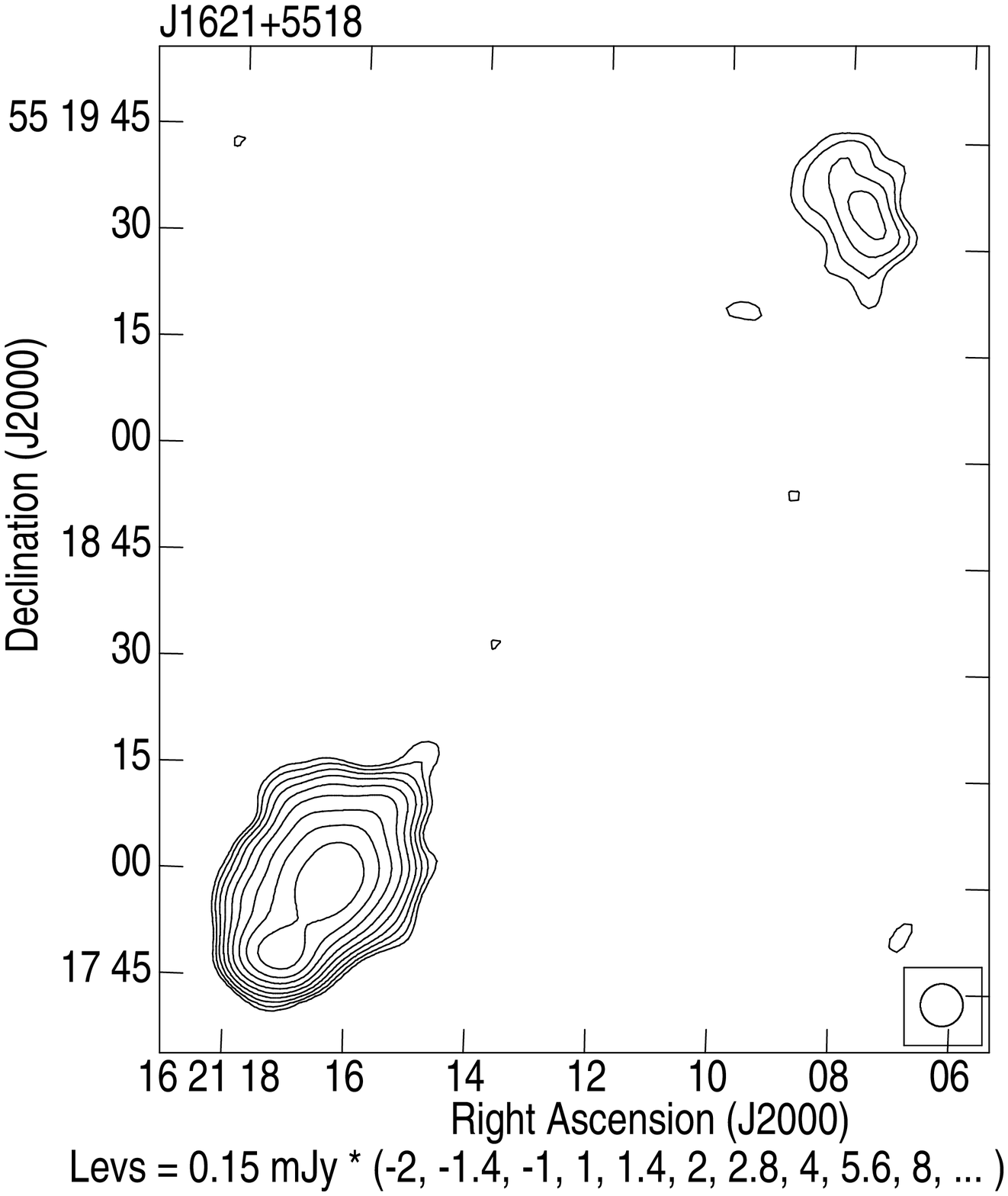} \\
\end{tabular}
\begin{tabular}{ll}
\vspace{0.95in}
\includegraphics[width=2.85in]{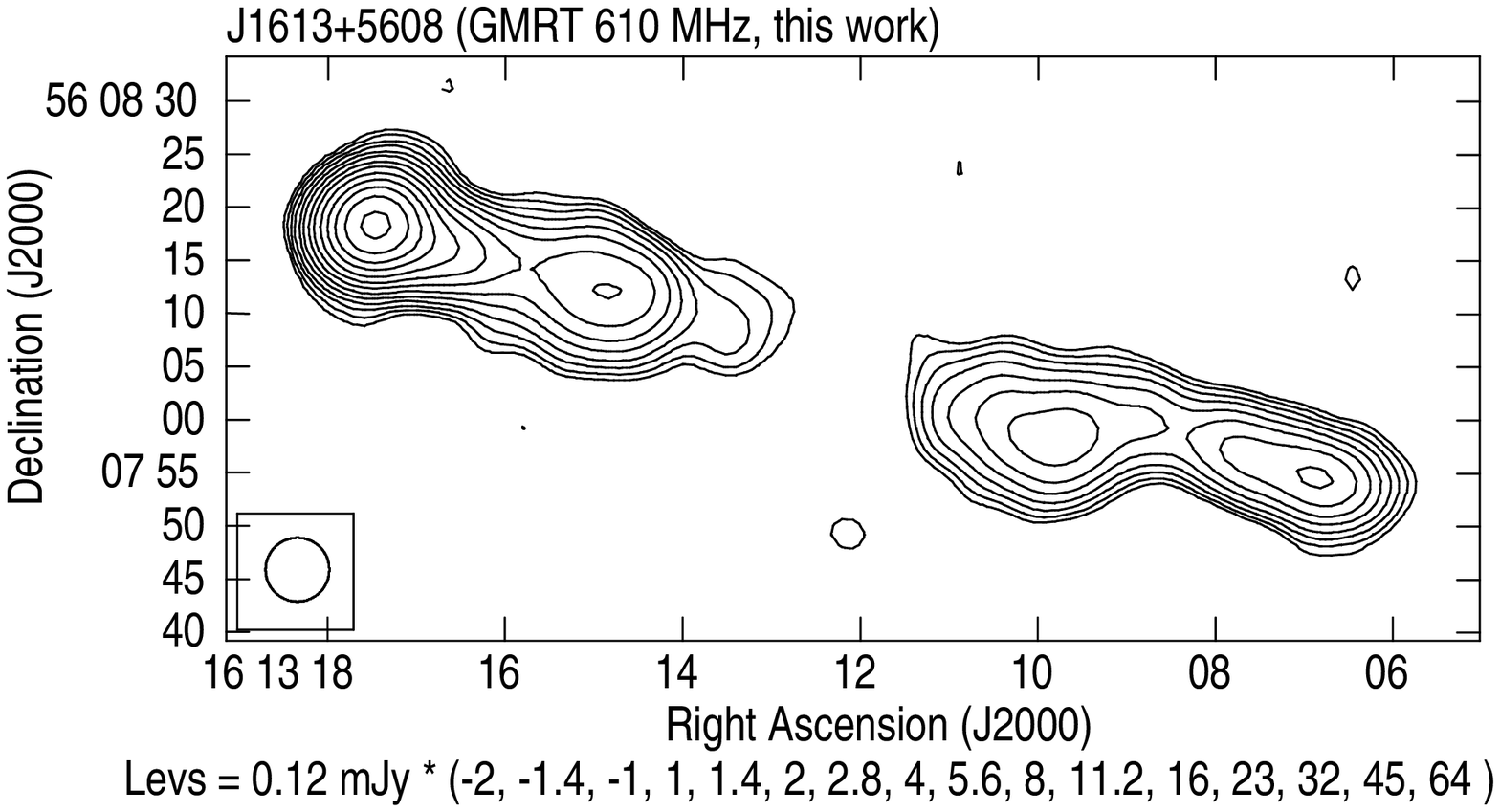}  &
\includegraphics[width=2.85in]{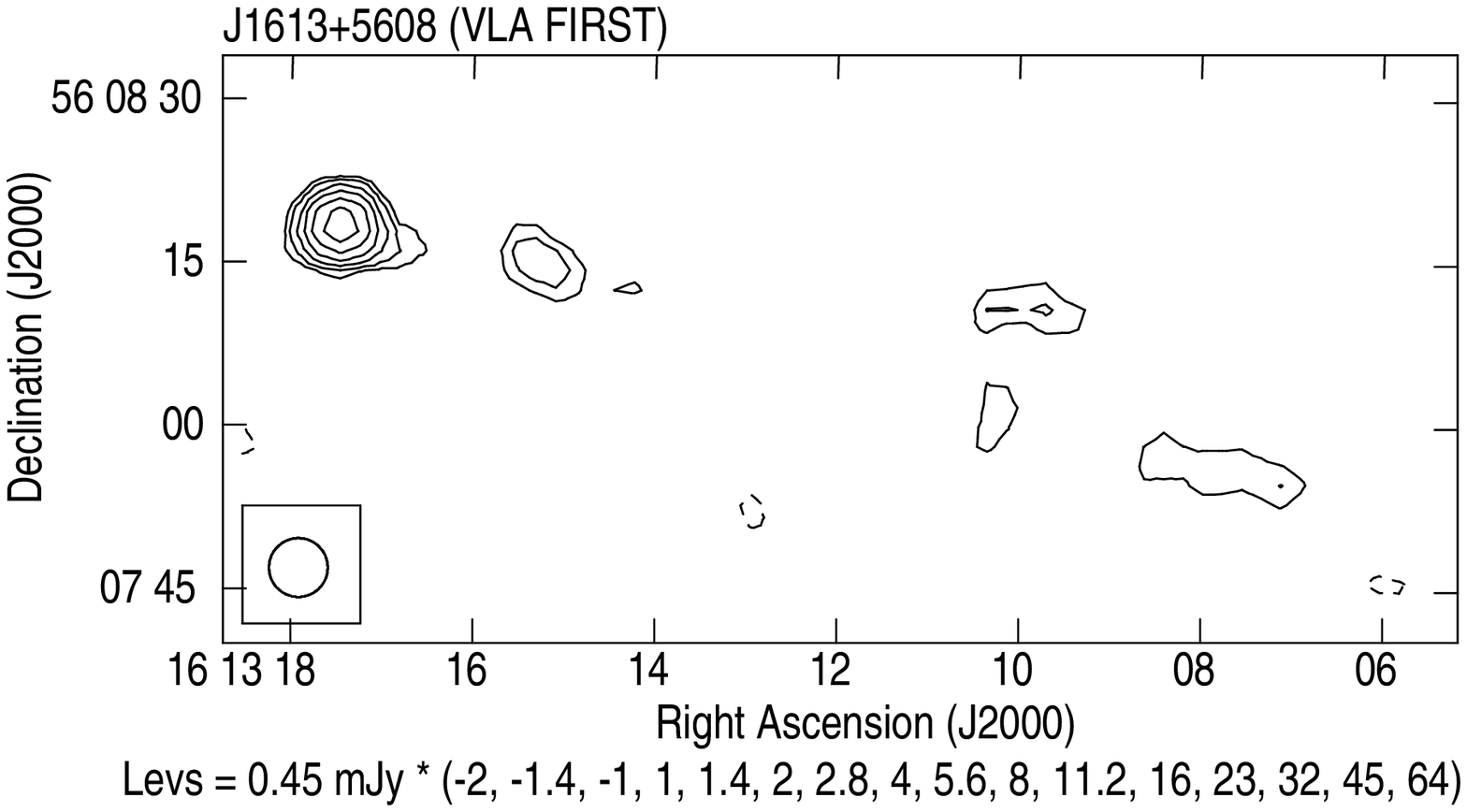} \\
\end{tabular}
\vspace{-0.95in}
\end{center}
\caption{A few radio sources with peculiar morphologies  and candidate relic AGNs. Top left: J1617+5432, top middle: J1611+5559, top right:J1621+5518.Bottom left: GMRT 610 MHz image of J1613+5608 and to right the VLA FIRST image of this source to highlight the non-detection of west-lobe at 1.4 GHz.} 
\label{fig:OTH1}
\end{figure*}

\subsubsection*{J1611$+$5432}

The spectroscopic redshift of this source (Figure \ref{fig:GRG1}, top panel, third from left) is 0.714 and for the angular size
2.2$'$, the linear size is 0.95 Mpc, hence we included this in the
category of GRS \ref{fig:GRG1}. Only the southern lobe is detected in VLA FIRST, but both lobes are detected in the NVSS with a flux density of 4 mJy for the southern lobe and the northern lobe at 2.9 mJy. The corresponding
values at 610 MHz are  11.9 mJy and 5.9 mJy respectively. The southern
lobe has a steeper spectral index of $-1.3$ as compared to the northern lobe
($-0.9$). 

\subsubsection*{J1608$+$5434}

The host galaxy of this GRS lies at spectroscopic redshift of 0.909. The angular size  1.8 arcmin of the source corresponds to a linear size of 0.85 Mpc Figure (\ref{fig:GRG1}, top panel, last). The core and southern lobe are blended in the NVSS and hence this appears as a double source in NVSS. Though the size is $<1$ Mpc, we have included this source here due to its higher redshift since there are very few GRS known in this redshift range.

\subsubsection*{J1609$+$5354}

This  giant radio quasar (Figure \ref{fig:GRG1}, bottom left panel) has a spectroscopic redshift of 0.99258. The
source is about 2.5$'$ in size which translates to a linear size of
1.1 Mpc. There are several interesting features for this giant radio
quasar. There are structures present in both the lobes, which suggests more than one hotspot. Recently \citet{2019MNRAS.482..240K} have studied a sample of powerful radio sources and modelled sources with such structures as due to precessing jets and speculated that such phenomena may be more common in radio loud AGNs. The other interesting aspect of this source is that the east lobe is resolved out in VLA FIRST (Figure \ref{fig:GRG1}, bottom middle panel), despite the fact
that both our 610 MHz image and VLA FIRST have  comparable angular resolution. The spectral index scaled equivalent rms noise at 1.4 GHz is $\sim 20$ $\mu$Jy beam$^{-1}$ for our 610 MHz images, which makes it much deeper than VLA FIRST, hence the clear detection of the east lobe with 610 MHz GMRT. This lobe is barely detected in the NVSS with flux density of 4.7 mJy. At 610 MHz the integrated flux density is 9.3 mJy, resulting in a spectral index of $-0.8$ for this lobe.

The west lobe is much brighter with flux density of 130 mJy at 610 MHz
and 58 mJy at 1.4 GHz from the NVSS. The spectral index is about $-1$,
steeper than that observed for lobes of other radio quasars. Such a
steeper spectral index for lobes of high redshift  radio sources could
also arise due to increased energy loss via the inverse-Compton
mechanism.

The  flux density of the core is 40.8 mJy at 610 MHz, 43.8 mJy at 1.4
GHz from VLA FIRST and 45.2 mJy at 1.4 GHz from the NVSS. The spectral index
of the core is thus flat, which is common for powerful radio quasars.
The total radio luminosity of this source at 1.4 GHz is $5.7 \times
10^{26}$ W~Hz$^{-1}$.

\subsubsection*{J1608$+$5611}

This is among the highest redshift giant radio quasars at a redshift of
1.32 (Figure \ref{fig:GRG1}, bottom right panel). The angular size is 2.6$'$ which corresponds to a linear size
of 1.3 Mpc. The armlength on the east is much longer than the west side resulting in a highly asymmetric axial ratio. The core has flat
spectrum. The total radio power is comparable to the GRS J1609$+$5354.

\begin{figure*}
\begin{center}
\begin{tabular}{ll}
\includegraphics[width=0.4\textwidth]{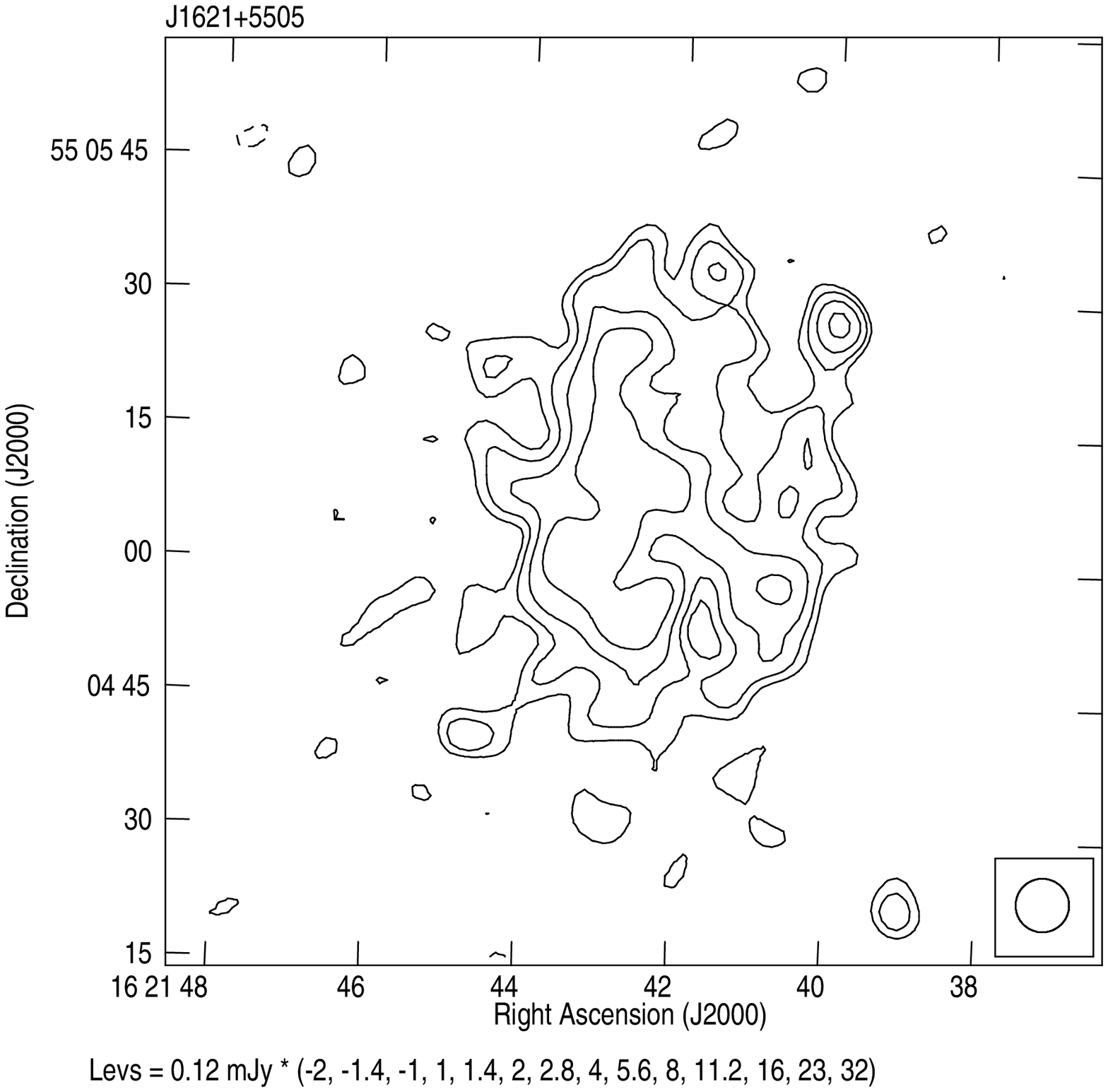} &
\includegraphics[width=0.44\textwidth]{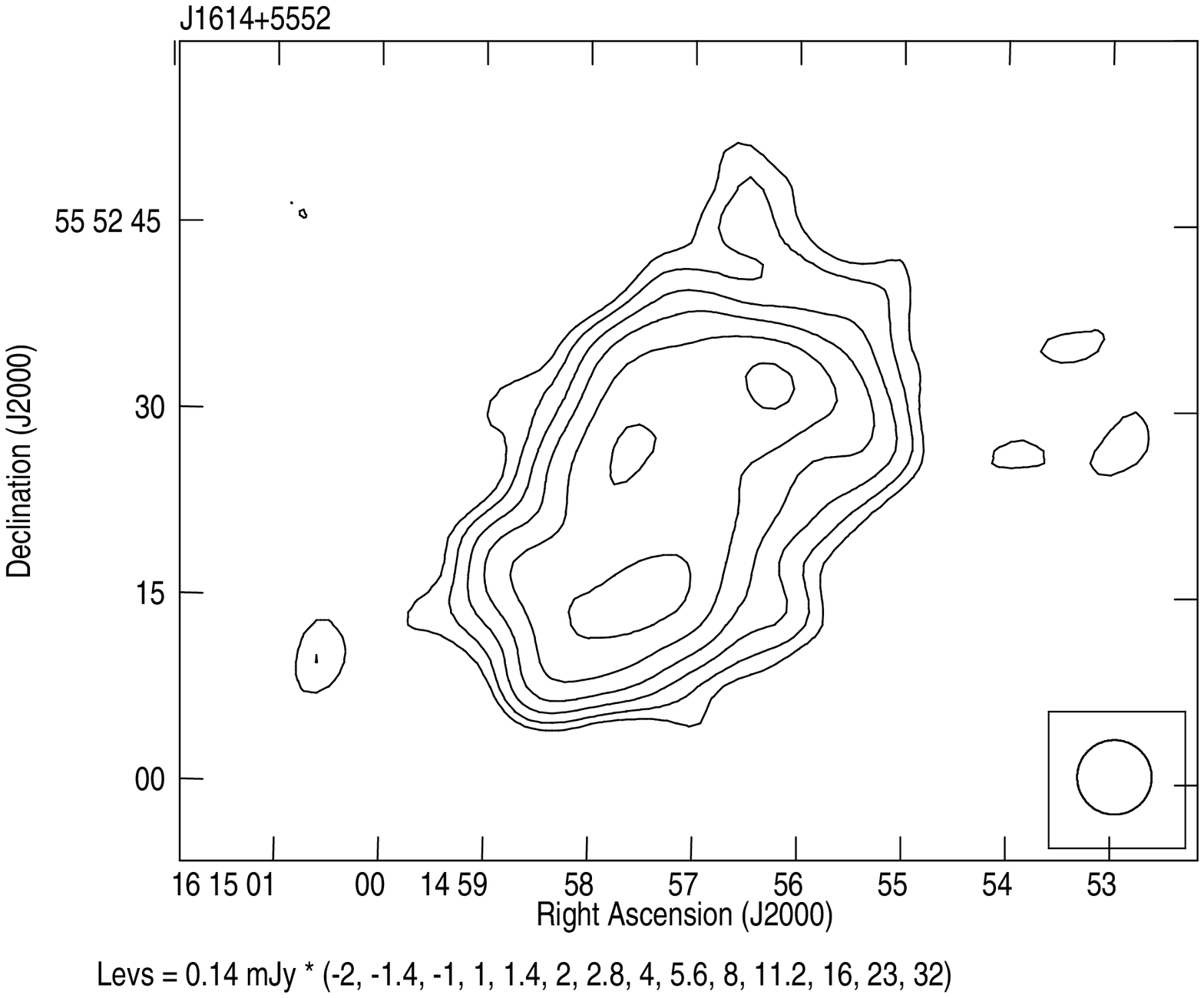}  \\
\includegraphics[width=0.44\textwidth]{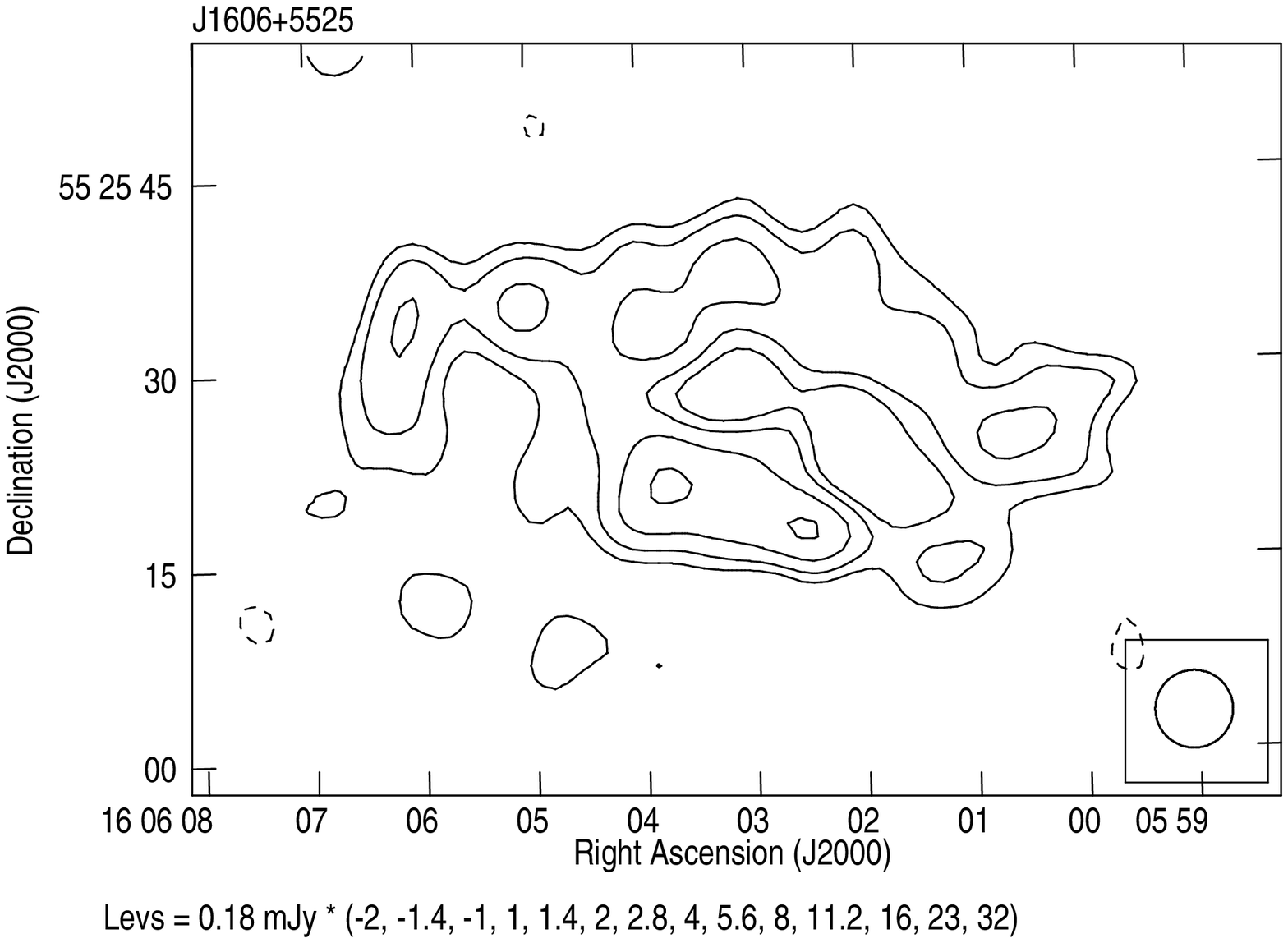} &
\includegraphics[width=0.44\textwidth]{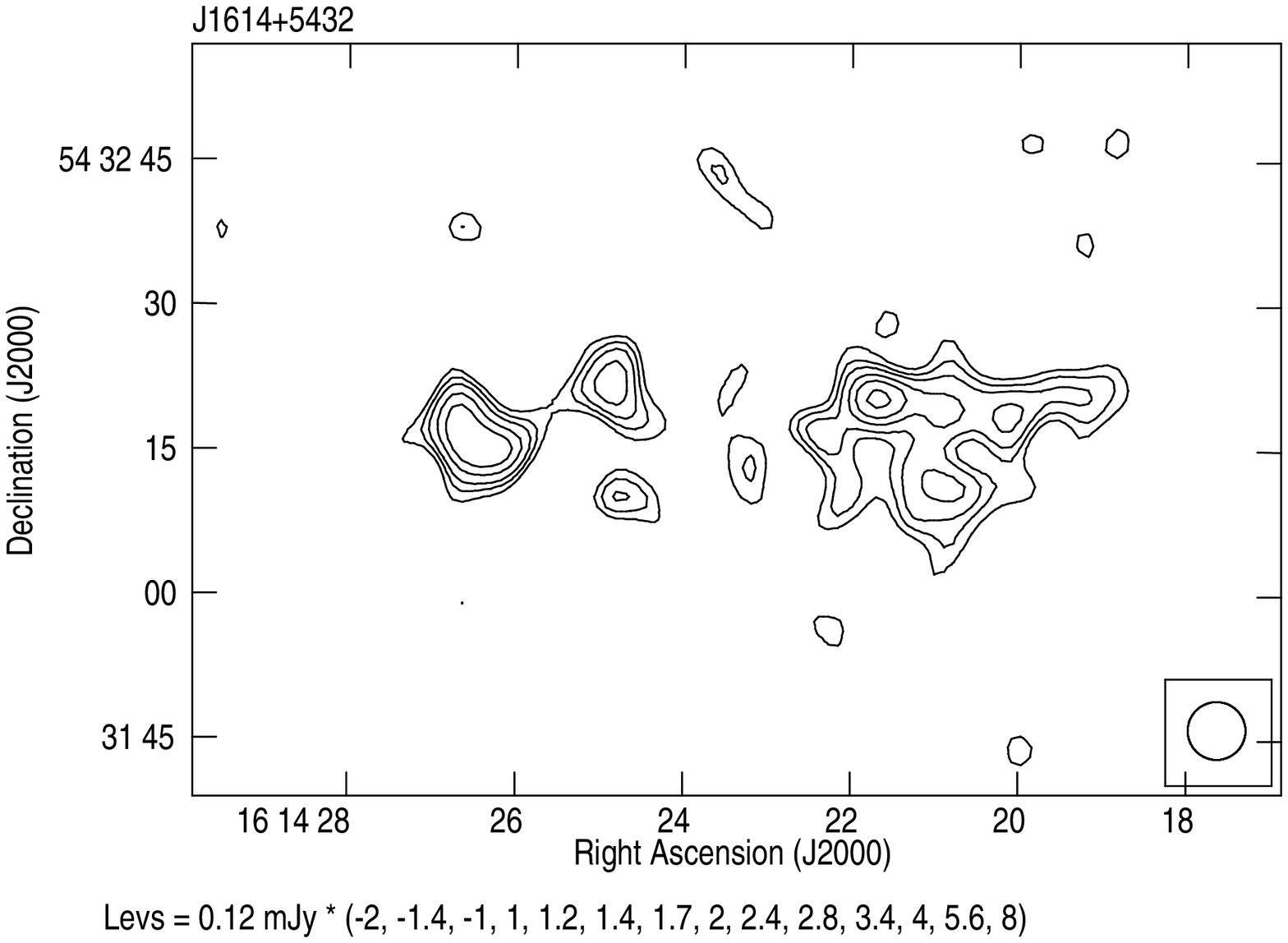} \\
\end{tabular}
\end{center}
\caption{Some nearby galaxies where extended radio emission is seen. Top Left: NGC 6143. Top right: CGCG 276-004. Bottom left: GALEXMSC J160603.68$+$552527.9. Bottom right:2MASX J16142202$+$5432180.}
\label{fig:SFG1}
\end{figure*}

\section{Other Peculiar sources}

In this section we discuss a few sources which are likely to be relic
sources or sources with very diffuse emission.  These sources have been selected by visually inspecting the image for extended morphology at 610
MHz. The purpose of this section is to highlight the possibility of
discovering peculiar sources and relics in the wide and deep radio
surveys.

\subsubsection*{J1617$+$5432}

This is a `compact' double--double radio source (Figure \ref{fig:OTH1}, top left panel). The optical counterpart
is located at the central region of the inner double and  has a 
spectroscopic redshift of 0.6999. The total angular size of the source
is less than an arcmin and hence the structure is not resolved in the NVSS.
The total flux density at 610 MHz is 10.7 mJy and at NVSS is 4.5 mJy
which corresponds to a spectral index of $-1.0$.   In the VLA FIRST survey,
only the inner compact hotspot pair is detected with total flux density
of 4 mJy. The radio morphology has two distinct emission regions, the
outer lobes are diffuse compared to compact inner lobes. The presence
of a clear compact hotspot pair in VLA FIRST further supports that this is
a re-started AGN. Since most of the double-double radio sources known
today are large angular size, the discovery of such relatively compact
sources with episodic activity may point to the scenario where the
signatures of episodic activity may be more common with shorter duty
cycle.

\subsubsection*{J1611$+$5559}

This is a very compact double  having an angular extent less than 0.5$'$
(Figure \ref{fig:OTH1}, top middle panel). The redshift of the galaxy is 0.314. The total flux density at
610 MHz is 15.7 mJy. The source is not catalogued in the VLA
FIRST.  We have inspected the VLA FIRST
image as well as our VLA image at 1.4 GHz with  marginally higher  resolution but with
comparable rms noise to VLA FIRST. A compact pair of hotspots separated
by about 6$''$, aligned in the north-east direction is seen. We have
estimated the flux around this using the {\tt AIPS} task {\tt TVSTAT}
which yielded a flux density of 3.6 mJy. The source is detected in the NVSS
with a flux density of 4.8 mJy which puts the spectral index between 610
MHz and 1.4 GHz at $-1.4$. This is a clear indication of presence of
relic emission.

\subsubsection*{J1621$+$5518}
The flux ratio between the two lobes  is  $>20$, making this one of the most
asymmetric doubles (Figure \ref{fig:OTH1}, top right panel).
A faint compact radio source is detected roughly
one-third distance to the north-west lobe and it has optical counterpart
in the SDSS with photometric redshift of 0.569. The flux density  of the
south-east lobe at 610 MHz is 25.4 mJy and at 1.4 GHz (NVSS) is 10.0
mJy. The flux density of north-west lobe is 2.1 mJy at 610  MHz and not
detected in our VLA image as well as VLA FIRST.

\subsubsection*{J1613$+$5608}

The radio morphology appears `normal' in the first look, but when
investigated in detail, this source has some interesting features
(Figure \ref{fig:OTH1}, bottom left panel). The
SDSS galaxy SDSS J161312.25$+$560800.5 with photometric redshift of 0.2 at
the `expected' location could be a possible core - there is no other
galaxy or quasar in the central region. The west lobe is resolved out in the
VLA FIRST survey (Figure \ref{fig:OTH1}, bottom right panel).
The total flux density at 610 MHz for the west lobe is
8.8 mJy and the NVSS flux density is 3.2\,mJy which gives a steep spectral
index of $-1.2$. For the east lobe the 610 MHz flux density is 18.4\,mJy and
at NVSS it is 6.5\,mJy which again gives a steep spectral index of
$-1.2$ for this lobe. Another interesting feature of this source is
knot like structures in the jet that are midway to the hot spot on either side.

\subsubsection*{J1615+5452}

This is a remnant radio AGN without any compact feature. Multi-frequency analysis of this AGN has been presented in  \citet{2020MNRAS.496.3381R}.

\section{Nearby galaxies}

Here we discuss a few regular nearby star forming galaxies showing extended morphology at 610 MHz.

\subsubsection*{NGC 6143}

NGC 6143 is a face on spiral galaxy at  a redshift of 0.0177 (Figure \ref{fig:SFG1}, top left panel). We have detected extended radio emission from this galaxy at 610 MHz. This source was not catalogued with PyBDSF due to its low surface brightness  and the smaller rms box size we adopted in PyBDSF. The
integrated flux density at 610 MHz is 13.2 mJy excluding the point source at the edge.  The source is detected in the NVSS with flux density of 6.2\,mJy. The point source is not detected in the VLA FIRST survey and in the high resolution VLA survey by this team, hence its contribution at 1.4 GHz is negligible. The spectral index of the source comes to be $-0.9 \pm 0.2$. The slightly steeper spectral index is consistent with the steeper spectral index observed for halo emission at low radio
frequencies in nearby galaxies such as NGC 891 \citep{1991A&A...246...10H}.

\subsubsection*{CGCG 276$-$004}

CGCG 276-004 is another face on spiral galaxy at redshift of 0.0529
which has been clearly detected in our observations (Figure \ref{fig:SFG1}, top right panel). This lies in a galaxy group \citep{2015AJ....149..171T}.  The integrated flux density is
12.8\,mJy at 610 MHz. The source has been detected in NVSS with a flux
density of 5.8\,mJy. The radio spectral between 610 MHz and 1.4 GHz is
$-0.95 \pm 0.2$ which is similar to  NGC 6143 discussed in the previous
section.

\subsubsection*{GALEXMSC J160603.68$+$552527.9}

This galaxy (Figure \ref{fig:SFG1}, bottom left panel) is at redshift of 0.0311 and SDSS classified it as `broad line galaxy'. This object has not been studied in detail  in the literature. The radio  emission is extremely diffuse and faint. The integrated flux density at 610 MHz is 10 mJy. The galaxy is detected in the NVSS with a flux density of 7.4 mJy.  The spectral index is much flatter than NGC 6143 and CGCG 276-004. This could be due to missing flux at 610 MHz because GALEXMSC J160603.68$+$552527.9 is close to a bright source hence the noise in this part of the image is more than a factor of two worse compared to rest of the image.

\subsubsection*{2MASX J16142202$+$5432180}

2MASX J16142202$+$5432180 is a galaxy at a redshift of 0.273 (Figure \ref{fig:SFG1},bottom right panel). The radio emission is very diffuse at 610 MHz with a total flux density of $\sim 5$ mJy.  The galaxy is not detected in the NVSS or VLA FIRST surveys. The radio luminosity at 610 MHz is $1.2 \times 10^{24}$ W~Hz$^{-1}$,  which suggests that the radio emission is  likely  due to an AGN. The star formation rate corresponding to this radio luminosity is about $\sim 40$ M$_{\odot}$  per year which is consistent with the star formation rate seen in Luminous Infra-red galaxies (LIRGs).

\section{Conclusions}

Here we have presented one of the deepest wide area GMRT 610 MHz surveys
of ELAIS\,N1 region covering 12.8 deg$^2$ with an rms noise of
$\sim 40$ $\mu$Jy beam$^{-1}$ at a resolution of 6 arcsec. 
This is equivalent to $\sim 20$ $\mu$Jy beam$^{-1}$ rms noise at 1.4 GHz for a spectral index of $-0.75$, which is several times deeper than VLA FIRST survey at similar resolution. This work provides valuable data for several multi-wavelength
studies of radio sources,  due to the wealth of ancillary data.
The main conclusions from this work are as follows.
\begin{itemize}
\item Above a threshold of 5$\sigma$, we have catalogued $\sim$ 6,400
radio sources. About  two-thirds of these are compact sources.
\item  Median spectral index between 610 MHz and 1.4 GHz is $-0.85
\pm 0.05$.
\item The radio source counts are among the deepest over a wide area. The source counts are consistent with the literature.
\item $\sim 90$ per cent of the sources have counterparts in optical or IR,
with majority having a photometric redshift.
\item Due to the improved sensitivity, we have discovered six giant radio
sources at redshift up to 1.3, three of them at redshift larger than 1,
suggesting that GRS are more common than proposed by earlier studies.
\item  We have found several extended sources  with steep spectra
which are candidate relic sources.
\item Extended emission from a few nearby galaxies was detected.
\end{itemize}

\section*{Acknowledgements}

We thank the staff of the GMRT that made these observations possible. GMRT is run by the National Centre for Radio Astrophysics of the Tata Institute of Fundamental Research. CHIC acknowledges the support of the Department of Atomic Energy, Government of India, under the project  12-R\&D-TFR-5.02-0700. This work was carried out using the data processing pipelines developed at the Inter-University Institute for Data Intensive Astronomy (IDIA). IDIA is a partnership of the University of Cape Town, the University of Pretoria and the University of the Western Cape. We acknowledge the use of the ilifu cloud computing facility - www.ilifu.ac.za, a partnership between the University of Cape Town, the University of the Western Cape, the University of Stellenbosch, Sol Plaatje University, the Cape Peninsula University of Technology and the South African Radio Astronomy Observatory. The ilifu facility is supported by contributions from the IDIA, the Computational Biology division at UCT and the Data Intensive Research Initiative of South Africa (DIRISA).
This research has made
use of the NASA/IPAC Extragalactic Database (NED), which is operated by
the Jet Propulsion Laboratory, California Institute of Technology, under
contract with the National Aeronautics and Space Administration.

\section*{Data availability}

The catalog of sources, presented in Table 1, is available as online supplementary material.

\bibliographystyle{mnras}
\bibliography{main}

\begin{thebibliography}{}
\makeatletter
\relax
\def\mn@urlcharsother{\let\do\@makeother \do\$\do\&\do\#\do\^\do\_\do\%\do\~}
\def\mn@doi{\begingroup\mn@urlcharsother \@ifnextchar [ {\mn@doi@}
  {\mn@doi@[]}}
\def\mn@doi@[#1]#2{\def\@tempa{#1}\ifx\@tempa\@empty \href
  {http://dx.doi.org/#2} {doi:#2}\else \href {http://dx.doi.org/#2} {#1}\fi
  \endgroup}
\def\mn@eprint#1#2{\mn@eprint@#1:#2::\@nil}
\def\mn@eprint@arXiv#1{\href {http://arxiv.org/abs/#1} {{\tt arXiv:#1}}}
\def\mn@eprint@dblp#1{\href {http://dblp.uni-trier.de/rec/bibtex/#1.xml}
  {dblp:#1}}
\def\mn@eprint@#1:#2:#3:#4\@nil{\def\@tempa {#1}\def\@tempb {#2}\def\@tempc
  {#3}\ifx \@tempc \@empty \let \@tempc \@tempb \let \@tempb \@tempa \fi \ifx
  \@tempb \@empty \def\@tempb {arXiv}\fi \@ifundefined
  {mn@eprint@\@tempb}{\@tempb:\@tempc}{\expandafter \expandafter \csname
  mn@eprint@\@tempb\endcsname \expandafter{\@tempc}}}

\bibitem[\protect\citeauthoryear{{Beck}, {Dobos}, {Budav{\'a}ri}, {Szalay}  \&
  {Csabai}}{{Beck} et~al.}{2016}]{Beck2016}
{Beck} R.,  {Dobos} L.,  {Budav{\'a}ri} T.,  {Szalay} A.~S.,   {Csabai} I.,
  2016, \mn@doi [\mnras] {10.1093/mnras/stw1009}, \href
  {http://adsabs.harvard.edu/abs/2016MNRAS.460.1371B} {460, 1371}

\bibitem[\protect\citeauthoryear{{Becker}, {White}  \& {Helfand}}{{Becker}
  et~al.}{1995}]{1995ApJ...450..559B}
{Becker} R.~H.,  {White} R.~L.,   {Helfand} D.~J.,  1995, \mn@doi [\apj]
  {10.1086/176166}, \href
  {https://ui.adsabs.harvard.edu/abs/1995ApJ...450..559B} {450, 559}

\bibitem[\protect\citeauthoryear{{Blumenthal} \& {Miley}}{{Blumenthal} \&
  {Miley}}{1979}]{1979A&A....80...13B}
{Blumenthal} G.,  {Miley} G.,  1979, \aap, \href
  {https://ui.adsabs.harvard.edu/abs/1979A&A....80...13B} {80, 13}

\bibitem[\protect\citeauthoryear{{Bonaldi}, {Bonato}, {Galluzzi}, {Harrison},
  {Massardi}, {Kay}, {De Zotti}  \& {Brown}}{{Bonaldi}
  et~al.}{2019}]{Bonaldi2019}
{Bonaldi} A.,  {Bonato} M.,  {Galluzzi} V.,  {Harrison} I.,  {Massardi} M.,
  {Kay} S.,  {De Zotti} G.,   {Brown} M.~L.,  2019, \mn@doi [\mnras]
  {10.1093/mnras/sty2603}, \href
  {https://ui.adsabs.harvard.edu/abs/2019MNRAS.482....2B} {482, 2}

\bibitem[\protect\citeauthoryear{{Bondi} et~al.,}{{Bondi}
  et~al.}{2007}]{2007A&A...463..519B}
{Bondi} M.,  et~al., 2007, \mn@doi [\aap] {10.1051/0004-6361:20066428}, \href
  {https://ui.adsabs.harvard.edu/abs/2007A&A...463..519B} {463, 519}

\bibitem[\protect\citeauthoryear{{Bonzini}, {Padovani}, {Mainieri},
  {Kellermann}, {Miller}, {Rosati}, {Tozzi}  \& {Vattakunnel}}{{Bonzini}
  et~al.}{2013}]{2013MNRAS.436.3759B}
{Bonzini} M.,  {Padovani} P.,  {Mainieri} V.,  {Kellermann} K.~I.,  {Miller}
  N.,  {Rosati} P.,  {Tozzi} P.,   {Vattakunnel} S.,  2013, \mn@doi [\mnras]
  {10.1093/mnras/stt1879}, \href
  {https://ui.adsabs.harvard.edu/abs/2013MNRAS.436.3759B} {436, 3759}

\bibitem[\protect\citeauthoryear{{Brienza} et~al.,}{{Brienza}
  et~al.}{2016}]{2016A&A...585A..29B}
{Brienza} M.,  et~al., 2016, \mn@doi [\aap] {10.1051/0004-6361/201526754},
  \href {https://ui.adsabs.harvard.edu/abs/2016A%26A...585A..29B} {585, A29}

\bibitem[\protect\citeauthoryear{{Callingham} et~al.,}{{Callingham}
  et~al.}{2017}]{2017ApJ...836..174C}
{Callingham} J.~R.,  et~al., 2017, \mn@doi [\apj]
  {10.3847/1538-4357/836/2/174}, \href
  {https://ui.adsabs.harvard.edu/abs/2017ApJ...836..174C} {836, 174}

\bibitem[\protect\citeauthoryear{{Chakraborty} et~al.,}{{Chakraborty}
  et~al.}{2019}]{2019MNRAS.490..243C}
{Chakraborty} A.,  et~al., 2019, \mn@doi [\mnras] {10.1093/mnras/stz2533},
  \href {https://ui.adsabs.harvard.edu/abs/2019MNRAS.490..243C} {490, 243}

\bibitem[\protect\citeauthoryear{{Condon}}{{Condon}}{1992}]{1992ARA&A..30..575C}
{Condon} J.~J.,  1992, \mn@doi [\araa] {10.1146/annurev.aa.30.090192.003043},
  \href {https://ui.adsabs.harvard.edu/abs/1992ARA%26A..30..575C} {30, 575}

\bibitem[\protect\citeauthoryear{{Condon}}{{Condon}}{2007}]{2007ASPC..380..189C}
{Condon} J.~J.,  2007, in {Afonso} J.,  {Ferguson} H.~C.,  {Mobasher} B.,
  {Norris} R.,  eds,  Astronomical Society of the Pacific Conference Series
  Vol. 380, Deepest Astronomical Surveys. p.~189

\bibitem[\protect\citeauthoryear{{Dabhade}, {Gaikwad}, {Bagchi},
  {Pandey-Pommier}, {Sankhyayan}  \& {Raychaudhury}}{{Dabhade}
  et~al.}{2017}]{2017MNRAS.469.2886D}
{Dabhade} P.,  {Gaikwad} M.,  {Bagchi} J.,  {Pandey-Pommier} M.,  {Sankhyayan}
  S.,   {Raychaudhury} S.,  2017, \mn@doi [\mnras] {10.1093/mnras/stx860},
  \href {https://ui.adsabs.harvard.edu/abs/2017MNRAS.469.2886D} {469, 2886}

\bibitem[\protect\citeauthoryear{{Dabhade} et~al.,}{{Dabhade}
  et~al.}{2020}]{2020A&A...635A...5D}
{Dabhade} P.,  et~al., 2020, \mn@doi [\aap] {10.1051/0004-6361/201935589},
  \href {https://ui.adsabs.harvard.edu/abs/2020A&A...635A...5D} {635, A5}

\bibitem[\protect\citeauthoryear{{Duncan} et~al.,}{{Duncan}
  et~al.}{2018a}]{Duncan2018a}
{Duncan} K.~J.,  et~al., 2018a, \mn@doi [\mnras] {10.1093/mnras/stx2536}, \href
  {https://ui.adsabs.harvard.edu/abs/2018MNRAS.473.2655D} {473, 2655}

\bibitem[\protect\citeauthoryear{{Duncan}, {Jarvis}, {Brown}  \&
  {R{\"o}ttgering}}{{Duncan} et~al.}{2018b}]{Duncan2018b}
{Duncan} K.~J.,  {Jarvis} M.~J.,  {Brown} M. J.~I.,   {R{\"o}ttgering} H.
  J.~A.,  2018b, \mn@doi [\mnras] {10.1093/mnras/sty940}, \href
  {https://ui.adsabs.harvard.edu/abs/2018MNRAS.477.5177D} {477, 5177}

\bibitem[\protect\citeauthoryear{{Eddington}}{{Eddington}}{1913}]{1913MNRAS..73..359E}
{Eddington} A.~S.,  1913, \mn@doi [\mnras] {10.1093/mnras/73.5.359}, \href
  {http://adsabs.harvard.edu/abs/1913MNRAS..73..359E} {73, 359}

\bibitem[\protect\citeauthoryear{{Eddington}}{{Eddington}}{1940}]{1940MNRAS.100..354E}
{Eddington} A.~S.,  1940, \mn@doi [\mnras] {10.1093/mnras/100.5.354}, \href
  {http://adsabs.harvard.edu/abs/1940MNRAS.100..354E} {100, 354}

\bibitem[\protect\citeauthoryear{{Fabian}}{{Fabian}}{2012}]{2012ARA&A..50..455F}
{Fabian} A.~C.,  2012, \mn@doi [\araa] {10.1146/annurev-astro-081811-125521},
  \href {https://ui.adsabs.harvard.edu/abs/2012ARA%26A..50..455F} {50, 455}

\bibitem[\protect\citeauthoryear{Garn, Green, Riley  \& Alexander}{Garn
  et~al.}{2008}]{2008MNRAS38375G}
Garn T.,  Green D.~A.,  Riley J.~M.,   Alexander P.,  2008, \mnras, 383, 75

\bibitem[\protect\citeauthoryear{{Hummel}, {Dahlem}, {van der Hulst}  \&
  {Sukumar}}{{Hummel} et~al.}{1991}]{1991A&A...246...10H}
{Hummel} E.,  {Dahlem} M.,  {van der Hulst} J.~M.,   {Sukumar} S.,  1991, \aap,
  \href {https://ui.adsabs.harvard.edu/abs/1991A%26A...246...10H} {246, 10}

\bibitem[\protect\citeauthoryear{{Intema}, {Jagannathan}, {Mooley}  \&
  {Frail}}{{Intema} et~al.}{2017}]{2017A&A...598A..78I}
{Intema} H.~T.,  {Jagannathan} P.,  {Mooley} K.~P.,   {Frail} D.~A.,  2017,
  \mn@doi [\aap] {10.1051/0004-6361/201628536}, \href
  {https://ui.adsabs.harvard.edu/abs/2017A&A...598A..78I} {598, A78}

\bibitem[\protect\citeauthoryear{{Ishwara-Chandra} \&
  {Saikia}}{{Ishwara-Chandra} \& {Saikia}}{1999}]{1999MNRAS.309..100I}
{Ishwara-Chandra} C.~H.,  {Saikia} D.~J.,  1999, \mn@doi [\mnras]
  {10.1046/j.1365-8711.1999.02835.x}, \href
  {https://ui.adsabs.harvard.edu/abs/1999MNRAS.309..100I} {309, 100}

\bibitem[\protect\citeauthoryear{{Ishwara-Chandra}, {Sirothia}, {Wadadekar},
  {Pal}  \& {Windhorst}}{{Ishwara-Chandra} et~al.}{2010}]{2010MNRAS.405..436I}
{Ishwara-Chandra} C.~H.,  {Sirothia} S.~K.,  {Wadadekar} Y.,  {Pal} S.,
  {Windhorst} R.,  2010, \mn@doi [\mnras] {10.1111/j.1365-2966.2010.16452.x},
  \href {https://ui.adsabs.harvard.edu/abs/2010MNRAS.405..436I} {405, 436}

\bibitem[\protect\citeauthoryear{{Kaiser}, {Dennett-Thorpe}  \&
  {Alexander}}{{Kaiser} et~al.}{1997}]{1997MNRAS.292..723K}
{Kaiser} C.~R.,  {Dennett-Thorpe} J.,   {Alexander} P.,  1997, \mn@doi [\mnras]
  {10.1093/mnras/292.3.723}, \href
  {https://ui.adsabs.harvard.edu/abs/1997MNRAS.292..723K} {292, 723}

\bibitem[\protect\citeauthoryear{{Kellermann}, {Sramek}, {Schmidt}, {Shaffer}
  \& {Green}}{{Kellermann} et~al.}{1989}]{1989AJ.....98.1195K}
{Kellermann} K.~I.,  {Sramek} R.,  {Schmidt} M.,  {Shaffer} D.~B.,   {Green}
  R.,  1989, \mn@doi [\aj] {10.1086/115207}, \href
  {https://ui.adsabs.harvard.edu/abs/1989AJ.....98.1195K} {98, 1195}

\bibitem[\protect\citeauthoryear{{Ker}, {Best}, {Rigby}, {R{\"o}ttgering}  \&
  {Gendre}}{{Ker} et~al.}{2012}]{2012MNRAS.420.2644K}
{Ker} L.~M.,  {Best} P.~N.,  {Rigby} E.~E.,  {R{\"o}ttgering} H.~J.~A.,
  {Gendre} M.~A.,  2012, \mn@doi [\mnras] {10.1111/j.1365-2966.2011.20235.x},
  \href {https://ui.adsabs.harvard.edu/abs/2012MNRAS.420.2644K} {420, 2644}

\bibitem[\protect\citeauthoryear{{Krause} et~al.,}{{Krause}
  et~al.}{2019}]{2019MNRAS.482..240K}
{Krause} M. G.~H.,  et~al., 2019, \mn@doi [\mnras] {10.1093/mnras/sty2558},
  \href {https://ui.adsabs.harvard.edu/abs/2019MNRAS.482..240K} {482, 240}

\bibitem[\protect\citeauthoryear{{Mahony} et~al.,}{{Mahony}
  et~al.}{2016}]{2016MNRAS.463.2997M}
{Mahony} E.~K.,  et~al., 2016, \mn@doi [\mnras] {10.1093/mnras/stw2225}, \href
  {https://ui.adsabs.harvard.edu/abs/2016MNRAS.463.2997M} {463, 2997}

\bibitem[\protect\citeauthoryear{{Ma{\l}ek} et~al.,}{{Ma{\l}ek}
  et~al.}{2018}]{2018A&A...620A..50M}
{Ma{\l}ek} K.,  et~al., 2018, \mn@doi [\aap] {10.1051/0004-6361/201833131},
  \href {https://ui.adsabs.harvard.edu/abs/2018A&A...620A..50M} {620, A50}

\bibitem[\protect\citeauthoryear{{Massardi}, {Bonaldi}, {Negrello},
  {Ricciardi}, {Raccanelli}  \& {de Zotti}}{{Massardi}
  et~al.}{2010}]{2010MNRAS.404..532M}
{Massardi} M.,  {Bonaldi} A.,  {Negrello} M.,  {Ricciardi} S.,  {Raccanelli}
  A.,   {de Zotti} G.,  2010, \mn@doi [\mnras]
  {10.1111/j.1365-2966.2010.16305.x}, \href
  {http://adsabs.harvard.edu/abs/2010MNRAS.404..532M} {404, 532}

\bibitem[\protect\citeauthoryear{{Mohan} \& {Rafferty}}{{Mohan} \&
  {Rafferty}}{2015}]{PyBDSF2015}
{Mohan} N.,  {Rafferty} D.,  2015, {PyBDSF: Python Blob Detection and Source
  Finder}, Astrophysics Source Code Library (\mn@eprint {ascl} {1502.007})

\bibitem[\protect\citeauthoryear{{Norris} et~al.,}{{Norris}
  et~al.}{2011}]{2011ApJ...736...55N}
{Norris} R.~P.,  et~al., 2011, \mn@doi [\apj] {10.1088/0004-637X/736/1/55},
  \href {https://ui.adsabs.harvard.edu/abs/2011ApJ...736...55N} {736, 55}

\bibitem[\protect\citeauthoryear{{Ocran}, {Taylor}, {Vaccari}  \&
  {Green}}{{Ocran} et~al.}{2017}]{Ocran2017}
{Ocran} E.~F.,  {Taylor} A.~R.,  {Vaccari} M.,   {Green} D.~A.,  2017, \mn@doi
  [\mnras] {10.1093/mnras/stx435}, \href
  {http://adsabs.harvard.edu/abs/2017MNRAS.468.1156O} {468, 1156}

\bibitem[\protect\citeauthoryear{{Ocran}, {Taylor}, {Vaccari}, {Ishwara-Chand
  ra}  \& {Prandoni}}{{Ocran} et~al.}{2020}]{Ocran2020}
{Ocran} E.~F.,  {Taylor} A.~R.,  {Vaccari} M.,  {Ishwara-Chand ra} C.~H.,
  {Prandoni} I.,  2020, \mn@doi [\mnras] {10.1093/mnras/stz2954}, \href
  {https://ui.adsabs.harvard.edu/abs/2020MNRAS.491.1127O} {491, 1127}

\bibitem[\protect\citeauthoryear{{Oliver} et~al.,}{{Oliver}
  et~al.}{2000}]{2000MNRAS.316..749O}
{Oliver} S.,  et~al., 2000, \mn@doi [\mnras]
  {10.1046/j.1365-8711.2000.03550.x}, \href
  {https://ui.adsabs.harvard.edu/abs/2000MNRAS.316..749O} {316, 749}

\bibitem[\protect\citeauthoryear{{Padovani}, {Miller}, {Kellermann},
  {Mainieri}, {Rosati}  \& {Tozzi}}{{Padovani}
  et~al.}{2011}]{2011ApJ...740...20P}
{Padovani} P.,  {Miller} N.,  {Kellermann} K.~I.,  {Mainieri} V.,  {Rosati} P.,
    {Tozzi} P.,  2011, \mn@doi [\apj] {10.1088/0004-637X/740/1/20}, \href
  {https://ui.adsabs.harvard.edu/abs/2011ApJ...740...20P} {740, 20}

\bibitem[\protect\citeauthoryear{{Perley} \& {Butler}}{{Perley} \&
  {Butler}}{2013}]{2013ApJS..204...19P}
{Perley} R.~A.,  {Butler} B.~J.,  2013, \mn@doi [\apjs]
  {10.1088/0067-0049/204/2/19}, \href
  {https://ui.adsabs.harvard.edu/abs/2013ApJS..204...19P} {204, 19}

\bibitem[\protect\citeauthoryear{{Pforr}, {Vaccari}, {Lacy}, {Maraston},
  {Nyland}, {Marchetti}  \& {Thomas}}{{Pforr} et~al.}{2019}]{Pforr2019}
{Pforr} J.,  {Vaccari} M.,  {Lacy} M.,  {Maraston} C.,  {Nyland} K.,
  {Marchetti} L.,   {Thomas} D.,  2019, \mn@doi [\mnras]
  {10.1093/mnras/sty3075}, \href
  {https://ui.adsabs.harvard.edu/abs/2019MNRAS.483.3168P} {483, 3168}

\bibitem[\protect\citeauthoryear{{Proctor}}{{Proctor}}{2016}]{2016ApJS..224...18P}
{Proctor} D.~D.,  2016, \mn@doi [\apjs] {10.3847/0067-0049/224/2/18}, \href
  {https://ui.adsabs.harvard.edu/abs/2016ApJS..224...18P} {224, 18}

\bibitem[\protect\citeauthoryear{{Randriamanakoto}, {Ishwara-Chandra}  \&
  {Taylor}}{{Randriamanakoto} et~al.}{2020}]{2020MNRAS.496.3381R}
{Randriamanakoto} Z.,  {Ishwara-Chandra} C.~H.,   {Taylor} A.~R.,  2020,
  \mn@doi [\mnras] {10.1093/mnras/staa1782}, \href
  {https://ui.adsabs.harvard.edu/abs/2020MNRAS.496.3381R} {496, 3381}

\bibitem[\protect\citeauthoryear{{Rowan-Robinson} et~al.,}{{Rowan-Robinson}
  et~al.}{2004}]{RowanRobinson2004}
{Rowan-Robinson} M.,  et~al., 2004, \mn@doi [\mnras]
  {10.1111/j.1365-2966.2004.07868.x}, \href
  {http://adsabs.harvard.edu/abs/2004MNRAS.351.1290R} {351, 1290}

\bibitem[\protect\citeauthoryear{{Rowan-Robinson} et~al.,}{{Rowan-Robinson}
  et~al.}{2008}]{RowanRobinson2008}
{Rowan-Robinson} M.,  et~al., 2008, \mn@doi [\mnras]
  {10.1111/j.1365-2966.2008.13109.x}, \href
  {https://ui.adsabs.harvard.edu/abs/2008MNRAS.386..697R} {386, 697}

\bibitem[\protect\citeauthoryear{{Rowan-Robinson}, {Gonzalez-Solares},
  {Vaccari}  \& {Marchetti}}{{Rowan-Robinson} et~al.}{2013}]{RowanRobinson2013}
{Rowan-Robinson} M.,  {Gonzalez-Solares} E.,  {Vaccari} M.,   {Marchetti} L.,
  2013, \mn@doi [\mnras] {10.1093/mnras/sts163}, \href
  {https://ui.adsabs.harvard.edu/abs/2013MNRAS.428.1958R} {428, 1958}

\bibitem[\protect\citeauthoryear{{Saxena} et~al.,}{{Saxena}
  et~al.}{2018}]{2018MNRAS.480.2733S}
{Saxena} A.,  et~al., 2018, \mn@doi [\mnras] {10.1093/mnras/sty1996}, \href
  {https://ui.adsabs.harvard.edu/abs/2018MNRAS.480.2733S} {480, 2733}

\bibitem[\protect\citeauthoryear{{Shirley} et~al.,}{{Shirley}
  et~al.}{2019}]{2019MNRAS.490..634S}
{Shirley} R.,  et~al., 2019, \mn@doi [\mnras] {10.1093/mnras/stz2509}, \href
  {https://ui.adsabs.harvard.edu/abs/2019MNRAS.490..634S} {490, 634}

\bibitem[\protect\citeauthoryear{{Sirothia}, {Dennefeld}, {Saikia}, {Dole},
  {Ricquebourg}  \& {Roland}}{{Sirothia} et~al.}{2009}]{2009MNRAS.395..269S}
{Sirothia} S.~K.,  {Dennefeld} M.,  {Saikia} D.~J.,  {Dole} H.,  {Ricquebourg}
  F.,   {Roland} J.,  2009, \mn@doi [\mnras]
  {10.1111/j.1365-2966.2009.14317.x}, \href
  {https://ui.adsabs.harvard.edu/abs/2009MNRAS.395..269S} {395, 269}

\bibitem[\protect\citeauthoryear{{Swarup}, {Ananthakrishnan}, {Kapahi}, {Rao},
  {Subrahmanya}  \& {Kulkarni}}{{Swarup} et~al.}{1991}]{1991CuSc...60...95S}
{Swarup} G.,  {Ananthakrishnan} S.,  {Kapahi} V.~K.,  {Rao} A.~P.,
  {Subrahmanya} C.~R.,   {Kulkarni} V.~K.,  1991, Current Science, \href
  {https://ui.adsabs.harvard.edu/abs/1991CuSc...60...95S} {60, 95}

\bibitem[\protect\citeauthoryear{{Tanaka} et~al.,}{{Tanaka}
  et~al.}{2018}]{Tanaka2018}
{Tanaka} M.,  et~al., 2018, \mn@doi [\pasj] {10.1093/pasj/psx077}, \href
  {https://ui.adsabs.harvard.edu/abs/2018PASJ...70S...9T} {70, S9}

\bibitem[\protect\citeauthoryear{{Tully}}{{Tully}}{2015}]{2015AJ....149..171T}
{Tully} R.~B.,  2015, \mn@doi [\aj] {10.1088/0004-6256/149/5/171}, \href
  {https://ui.adsabs.harvard.edu/abs/2015AJ....149..171T} {149, 171}

\bibitem[\protect\citeauthoryear{{Vaccari}}{{Vaccari}}{2015}]{Vaccari2015}
{Vaccari} M.,  2015, in The Many Facets of Extragalactic Radio Surveys: Towards
  New Scientific Challenges. p.~27 (\mn@eprint {arXiv} {1604.02353})

\bibitem[\protect\citeauthoryear{{Vaccari}}{{Vaccari}}{2016}]{Vaccari2016}
{Vaccari} M.,  2016, \mn@doi [The Universe of Digital Sky Surveys]
  {10.1007/978-3-319-19330-4_10}, \href
  {http://adsabs.harvard.edu/abs/2016ASSP...42...71V} {42, 71}

\bibitem[\protect\citeauthoryear{{Vaccari} et~al.,}{{Vaccari}
  et~al.}{2005}]{Vaccari2005}
{Vaccari} M.,  et~al., 2005, \mn@doi [\mnras]
  {10.1111/j.1365-2966.2005.08773.x}, \href
  {http://adsabs.harvard.edu/abs/2005MNRAS.358..397V} {358, 397}

\bibitem[\protect\citeauthoryear{{Vaccari} et~al.,}{{Vaccari}
  et~al.}{2010}]{Vaccari2010}
{Vaccari} M.,  et~al., 2010, \mn@doi [\aap] {10.1051/0004-6361/201014694},
  \href {http://adsabs.harvard.edu/abs/2010A%26A...518L..20V} {518, L20}

\bibitem[\protect\citeauthoryear{{Whittam} et~al.,}{{Whittam}
  et~al.}{2013}]{2013MNRAS.429.2080W}
{Whittam} I.~H.,  et~al., 2013, \mn@doi [\mnras] {10.1093/mnras/sts478}, \href
  {https://ui.adsabs.harvard.edu/abs/2013MNRAS.429.2080W} {429, 2080}

\bibitem[\protect\citeauthoryear{{Wilman} et~al.,}{{Wilman}
  et~al.}{2008}]{Wilman2008}
{Wilman} R.~J.,  et~al., 2008, \mn@doi [\mnras]
  {10.1111/j.1365-2966.2008.13486.x}, \href
  {https://ui.adsabs.harvard.edu/abs/2008MNRAS.388.1335W} {388, 1335}

\bibitem[\protect\citeauthoryear{{Windhorst}, {Miley}, {Owen}, {Kron}  \&
  {Koo}}{{Windhorst} et~al.}{1985}]{1985ApJ...289..494W}
{Windhorst} R.~A.,  {Miley} G.~K.,  {Owen} F.~N.,  {Kron} R.~G.,   {Koo} D.~C.,
   1985, \mn@doi [\apj] {10.1086/162911}, \href
  {https://ui.adsabs.harvard.edu/abs/1985ApJ...289..494W} {289, 494}

\makeatother
\end{thebibliography}

\bsp

\label{lastpage}

\end{document}